\documentclass[a4paper]{aa}

\usepackage{graphicx,natbib}
\usepackage{txfonts}
\usepackage{color}
\usepackage[usenames,dvipsnames]{xcolor}

\newcommand{\bz}{$\langle B_\mathrm{z} \rangle$}
\newcommand{\vir}{CU\,Vir}
\newcommand{\teff}{$T_{\rm eff}$}
\newcommand{\lgg}{$\log g$}
\newcommand{\vs}{$v_{\rm e}\sin i$}
\newcommand{\kms}{km\,s$^{-1}$}

\newcommand{\figps}[1]{\resizebox{\hsize}{!}{\rotatebox{0}{\includegraphics{#1}}}}

\newcommand{\fifps}[2]{\centering\resizebox{#1}{!}{\includegraphics{#2}}}
\newcommand{\firrps}[2]{\resizebox{#1}{!}{\rotatebox{270}{\includegraphics{#2}}}}

\newcommand{\beq}{\begin{equation}}
\newcommand{\eeq}{\end{equation}}

\begin{document}
\title{Magnetic field topology of the unique chemically \\ peculiar star CU Virginis\thanks{Based on observations obtained at the Telescope Bernard Lyot (USR5026) operated by the Observatoire Midi-Pyr\'en\'ees, Universit\'e de Toulouse (Paul Sabatier), Centre National de la Recherche Scientifique of France}}

\author{O.~Kochukhov\inst{1}
   \and T.~L\"uftinger\inst{2}
   \and C.~Neiner\inst{3}
   \and E.~Alecian\inst{4,3}
   \and the MiMeS collaboration}

\institute{
Department of Physics and Astronomy, Uppsala University, Box 516, SE-75120 Uppsala, Sweden
\and
Institut f\"ur Astrophysik, Universit\"at Wien, T\"urkenschanzstr. 17, A-1180 Wien, Austria
\and
LESIA, Observatoire de Paris, CNRS UMR 8109, UPMC, Universit\'e Paris Diderot, 5 place Jules Janssen, 92190 Meudon, France
\and
UJF-Grenoble 1 / CNRS-INSU, Institut de Plan\'etologie et d'Astrophysique de Grenoble (IPAG) UMR 5274, Grenoble, F-38041, France
}

\date{Received 21 January 2014 / Accepted xx April 2014}

\titlerunning{Magnetic field topology of CU Virginis}
\authorrunning{O.~Kochukhov et al.}

\abstract
{
The late-B magnetic chemically peculiar star  \vir\ is one of the fastest rotators among the intermediate-mass stars with strong fossil magnetic fields. It shows a prominent rotational modulation of the spectral energy distribution and absorption line profiles due to chemical spots and exhibits a unique strongly beamed variable radio emission.
}
{
Little is known about the magnetic field topology of \vir. In this study we aim to derive, for the first time, detailed maps of the magnetic field distribution over the surface of this star.
}
{
We use high-resolution spectropolarimetric observations covering the entire rotational period. These data are interpreted using a multi-line technique of least-squares deconvolution (LSD) and a new Zeeman Doppler imaging code based on detailed polarised radiative transfer modelling of the Stokes $I$ and $V$ LSD profiles. This new magnetic inversion approach relies on the spectrum synthesis calculations over the full wavelength range covered by observations and does not assume that the LSD profiles behave as a single spectral line with mean parameters.
}
{
We present magnetic and chemical abundance maps derived from the Si and Fe lines. Mean polarisation profiles of both elements reveal a significant departure of the magnetic field topology of \vir\ from the commonly assumed axisymmetric dipolar configuration. The field of \vir\ is  dipolar-like, but clearly non-axisymmetric, showing a large difference of the field strength between the regions of opposite polarity. The main relative abundance depletion features in both Si and Fe maps coincide with the weak-field region in the magnetic map.
}
{
Detailed information on the distorted dipolar magnetic field topology of \vir\ provided by our study is essential for understanding chemical spot formation, radio emission, and rotational period variation of this star.
}


\maketitle

\section{Introduction}
\label{intro}

The bright late-B star \vir\ (HR\,5313, HD\,124224, HIP\,69389) is one of the best known intermediate-mass magnetic chemically peculiar (CP) stars. This object shows abnormally strong lines of Si, weak lines of He, and a notable spectroscopic \citep{deutsch:1952} and photometric \citep{hardie:1958} variability with a period of $\sim$\,0.5~d. These periodic changes arise due to a combination of stellar rotation and an inhomogeneous distribution of chemical elements over the stellar surface. It is generally believed that high-contrast chemical spots are produced by an anisotropic atomic diffusion in the presence of strong, globally-organised magnetic field \citep{michaud:1981,alecian:1981,alecian:2010}, although a detailed diffusion theory capable of explaining the surface structure of individual magnetic CP stars is still lacking.

Being one of the most rapidly rotating magnetic CP stars, \vir\ represents an optimal target for the reconstruction of the surface chemical spot distributions with the Doppler imaging (DI) method. Using this technique, maps of He, Mg, Si, Cr, and Fe were obtained for \vir\ by different authors \citep{goncharskii:1983,hiesberger:1995,hatzes:1997,kuschnig:1999}. The study by \citet{krticka:2012} showed that these uneven chemical distributions provide a successful quantitative explanation for the observed spectrophotometric behaviour of \vir\ in the ultraviolet and optical wavelength regions.

The fast rotation and stability of the photometric light curves of \vir\ enable a very precise period determination and allows one to study subtle long-term variations of stellar rotation. \vir\ is one of a few stars for which rotational period changes have been securely detected. Remarkably, this star appears to show period glitches \citep{pyper:1998,pyper:2013} or possibly cyclic period variation \citep{mikulasek:2011} rather than a constantly increasing period as expected from magnetic spin down models \citep{ud-doula:2009}. The origin of these period changes is currently unknown.

In addition to the photometric and spectroscopic variability typical of magnetic CP stars, \vir\ exhibits a unique radio emission signature, not seen in any other star. \citet{trigilio:2000} reported two sharp 100 per cent circularly polarised radio pulses occurring at particular rotational phases. This emission was subsequently studied in different radio bands \citep{trigilio:2008,trigilio:2011,ravi:2010,lo:2012} and interpreted as an electron cyclotron maser emission originating in the stellar magnetosphere above one of the magnetic poles \citep[see][and references therein]{lo:2012}. The sharp radio emission pulses can also be used for a very precise measurement of the stellar rotational period. However, it is not clear if the phase shifts observed in the radio are related to intrinsic variation of the stellar rotation or to instabilities in the region where the radio emission is produced \citep{ravi:2010,pyper:2013}.

Evidently, the magnetic field is a key parameter for understanding different variability phenomena observed in \vir. The strength and geometry of the surface magnetic field influences the photospheric distribution of chemical elements, while the structure of the magnetosphere at a distance of 2--3 stellar radii is the main ingredient of any possible non-thermal radio emission process. The presence of a kG-strength magnetic field in \vir\ was established by \citet{landstreet:1977} and \citet{borra:1980} using a Balmer line magnetometer. Their mean longitudinal magnetic field (\bz) measurements revealed a smooth, roughly sinusoidal variation with an amplitude of $\approx$\,600~G and reversing sign. Several oblique dipolar models were derived using these \bz\ data \citep{borra:1980,trigilio:2000}, resulting in estimates of the polar field strength $B_{\rm d}$\,=\,3.0--4.5~kG and magnetic obliquity $\beta$\,=\,70--90\degr, depending on the assumed inclination angle $i$. On the other hand, \citet{hatzes:1997} and \citet{glagolevskij:2002} argued that the magnetic field geometry of \vir\ is better described by an offset dipole or by a combination of dipolar and quadrupolar components of comparable strengths. Observations of the mean longitudinal magnetic field available in the literature are insufficient to definitively distinguish between these alternative magnetic field models.

The goal of our investigation is to obtain a much more precise and detailed picture of the magnetic field topology of \vir\ by using new high-resolution circular polarisation observations. To interpret these data we developed a new version of the Zeeman Doppler imaging (ZDI) methodology, which for the first time combines detailed polarised spectrum synthesis calculations  based on realistic model atmospheres with a multi-line approach widely used in stellar spectropolarimetry.

The rest of this paper is organised as follows. Sect.~\ref{methods} presents spectropolarimetric observations and outlines corresponding analysis methodology, including calculation of the mean polarisation profiles and their interpretation in terms of the magnetic and chemical stellar surface maps. In Sect.~\ref{results} we present revised stellar parameters, discuss new longitudinal field measurements, and derive magnetic field topology and chemical abundance distributions with the help of ZDI inversions. The results of our investigation are summarised and discussed in the context of other recent studies in Sect.~\ref{discussion}.

\section{Methods}
\label{methods}

\subsection{Spectropolarimetric observations}

We collected 24 spectropolarimetric observations of \vir\ in the context of the Magnetism in Massive Stars (MiMeS) large program. The Stokes $I$ and $V$ spectra were obtained over a time span of 26 months in 2009--2011 with the Narval spectropolarimeter attached to the 2-m T\'elescope Bernard Lyot at the Pic du Midi observatory. Narval is a fibre-fed cross-dispersed echelle spectropolarimeter, covering the wavelength range from 3694 to 10483~\AA\ at the resolution of $\lambda/\Delta\lambda=65\,000$.

The spectra of \vir\ were processed with the {\sc Libre-Esprit} reduction package \citep{donati:1997}. The extracted echelle orders were subsequently renormalised to the continuum and merged using a set of dedicated IDL routines. The signal-to-noise (S/N) ratio of the resulting spectra varied between 700 and 1500 per extracted pixel, as measured at $\lambda=5400$~\AA.

Each Narval observation of \vir\ consisted of a sequence of four sub-exposures obtained at two different orientations of the quarter wave retarder plate relative to the beam splitter. Combining the resulting left- and right-hand polarised spectra as discussed by \citet{donati:1997} and \citet{bagnulo:2009} allows one to minimise spurious polarisation and to derive a diagnostic null spectrum, which is useful for identifying  residual instrumental artefacts.

Rotational phases of all observations of \vir\ analysed in our study were evaluated with the help of the variable-period ephemeris suggested by \citet{mikulasek:2011}. This allowed us to phase the Narval data with the earlier magnetic field measurements by \citet{borra:1980}. We note that the period variability discussed by \citet{mikulasek:2011} and in other recent publications has a negligible impact on the relative phases of the Narval spectra since our observations were collected during a relatively short period. The mean value of the rotational period during Narval observations, 0.5207130~d, is also identical within error bars to the constant period of $0.5207137\pm0.0000010$~d proposed by \citet{pyper:2013} for the time period 1993--2012.

Information on the observing dates, heliocentric Julian dates, rotational phases, and S/N ratio of individual Narval observations of \vir\ is provided in the first four columns of Table~\ref{tbl:obs}.

\begin{table*}[!th]
\centering
\caption{Journal of spectropolarimetric observations of \vir. 
\label{tbl:obs}}
\begin{tabular}{cccccr}
\hline\hline
mid-UT date & mid-HJD & $\varphi$ & S/N & S/N$_{\rm LSD}$ & \bz\ (G)~~ \\
\hline
2009-03-10 &  2454900.6241 & 0.4697 &  1400 & 22700 & $-274\pm~~39$ \\
2009-03-12 &  2454902.5167 & 0.1043 &  1410 & 24000 &  $758\pm~~61$ \\
2009-03-12 &  2454903.4882 & 0.9700 &   860 & 14700 &  $531\pm105$ \\
2009-03-13 &  2454904.4845 & 0.8832 &  1460 & 24700 &  $362\pm~~56$ \\
2009-03-15 &  2454905.5541 & 0.9374 &  1480 & 25100 &  $506\pm~~60$ \\
2009-03-15 &  2454906.4955 & 0.7452 &  1110 & 19100 & $-267\pm~~52$ \\
2009-03-16 &  2454907.4898 & 0.6548 &  1320 & 22300 & $-519\pm~~40$ \\
2009-03-18 &  2454908.5223 & 0.6376 &  1390 & 23600 & $-585\pm~~38$ \\
2009-03-19 &  2454909.5020 & 0.5192 &  1200 & 20000 & $-517\pm~~45$ \\
2009-04-13 &  2454935.4881 & 0.4239 &  1400 & 22000 &  $-30\pm~~41$ \\
2009-04-14 &  2454935.5242 & 0.4931 &  1470 & 23600 & $-458\pm~~37$ \\
2009-04-23 &  2454945.4871 & 0.6262 &   930 & 16000 & $-681\pm~~55$ \\
2010-03-29 &  2455284.5805 & 0.8355 &   660 & 10500 &  $-97\pm117$ \\
2010-04-06 &  2455292.5987 & 0.2341 &   970 & 16600 &  $901\pm~~66$ \\
2010-04-10 &  2455296.5679 & 0.8566 &  1280 & 21200 &  $259\pm~~61$ \\
2010-04-20 &  2455306.5932 & 0.1097 &  1330 & 22700 &  $711\pm~~64$ \\
2011-01-15 &  2455576.7393 & 0.9108 &  1370 & 23800 &  $578\pm~~61$ \\
2011-01-17 &  2455578.7462 & 0.7649 &  1000 &  8700 & $-351\pm124$ \\
2011-02-04 &  2455596.7326 & 0.3067 &   890 & 12400 &  $563\pm~~79$ \\
2011-03-19 &  2455639.6436 & 0.7151 &  1380 & 22800 & $-439\pm~~42$ \\
2011-03-20 &  2455640.5735 & 0.5011 &  1360 & 22100 & $-574\pm~~40$ \\
2011-04-05 &  2455656.5808 & 0.2422 &  1060 & 17900 & $707\pm~~60$ \\
2011-04-06 &  2455657.5109 & 0.0284 &  1030 & 17500 &  $942\pm~~89$ \\
2011-05-08 &  2455690.4711 & 0.3269 &  1160 & 18800 &  $505\pm~~50$ \\
\hline
\end{tabular}
\tablefoot{The columns give the UT date of mid-observation, the corresponding Heliocentric Julian date, rotational phase, the signal-to-noise ratio per pixel measured at $\lambda=5400$~\AA, the signal-to-noise ratio per 3~\kms\ velocity bin in the Stokes $V$ LSD profiles, and the mean longitudinal magnetic field determined from these LSD profiles.}
\end{table*}

\subsection{Multi-line polarisation analysis}

Owing to a large rotational Doppler broadening of the lines in the spectrum of \vir\ and to a moderate strength of its magnetic field, even with our very high quality spectropolarimetric observations the circular polarisation signatures can be detected at the significance level of 3--5$\sigma$ only in a few very strong metal lines and for a limited range of rotational phases. To enhance the analysis of the Stokes $V$ spectra we applied a multi-line technique of least-squares deconvolution \citep[LSD,][]{donati:1997,kochukhov:2010a}, combining hundreds of spectral lines into high S/N ratio mean profiles. 

The line mask necessary for application of LSD was obtained from the VALD data base \citep{piskunov:1995,kupka:1999}, using the stellar parameters discussed below (see Sect.~\ref{params}). In total, we used 359 spectral lines with a central depth (before instrumental and rotational broadening) of $\ge$\,10 per cent of the continuum. The lines were chosen in the 4000--6865~\AA\ wavelength interval, avoiding the regions affected by the broad hydrogen Balmer line wings. Three sets of the Stokes $I$ and $V$ profiles were calculated: the one based on all metal lines, the mean Si profile (based on 63 lines), and the mean Fe profile (based on 202 lines). The LSD profiles derived from the full line mask were used for the longitudinal field measurements, while the mean Si and Fe profiles were employed for the magnetic inversions. All LSD profiles were calculated within $\pm350$~\kms\ of the laboratory line positions, using 3~\kms\ velocity bins. 

\begin{figure}[!th]
\centering
\figps{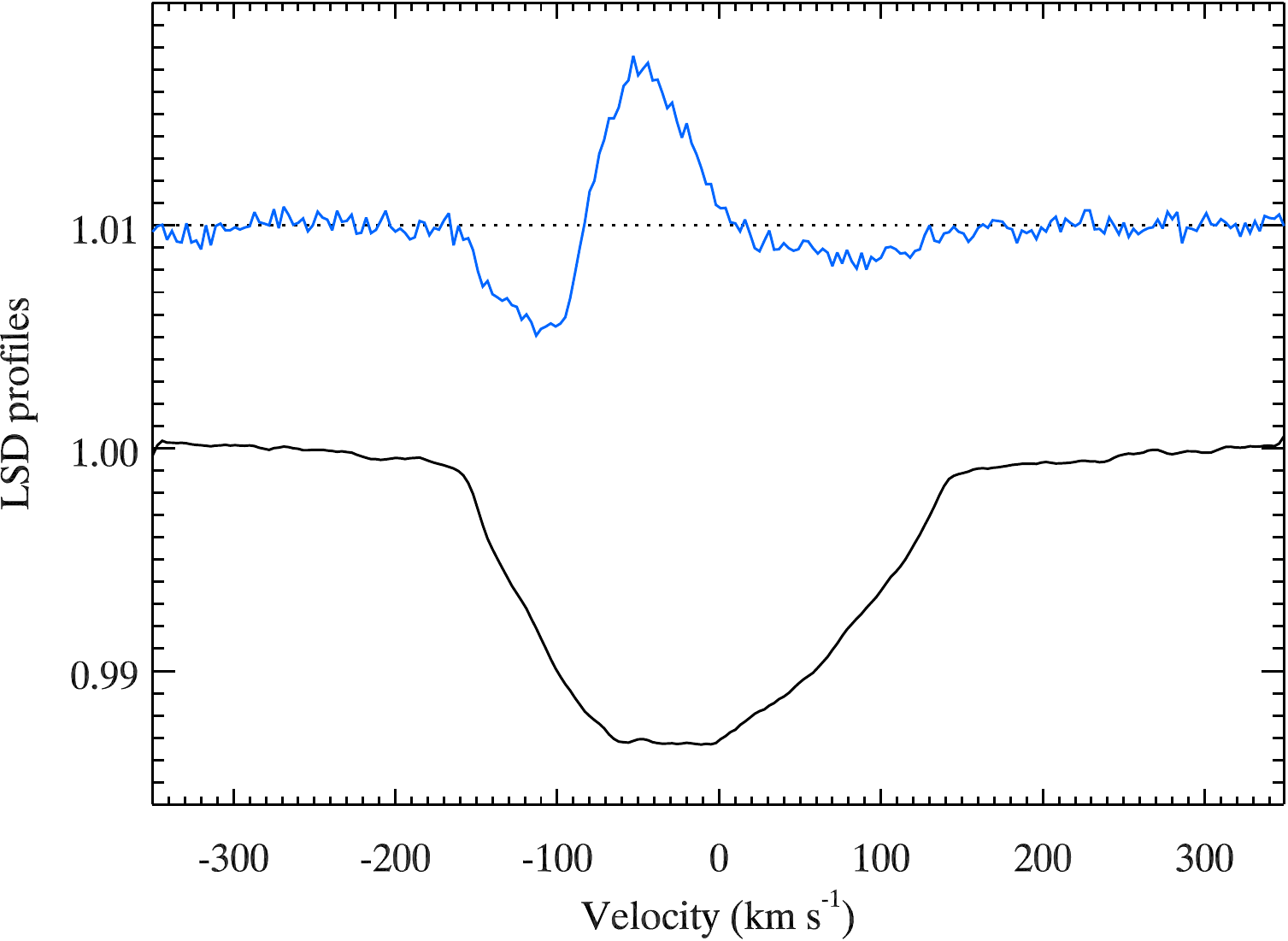}
\caption{LSD Stokes $I$ (bottom) and Stokes $V$ (top) profiles of \vir\ derived from the observation obtained on Apr 13, 2009 using the full metal line mask. The Stokes $V$ profile is offset vertically and expanded by a factor of 10 relative to Stokes $I$.}
\label{fig:lsd}
\end{figure}

The LSD profiles for specific elements were recovered using the multi-profile version of the LSD method discussed by \citet{kochukhov:2010a}. This means that, in practice, we calculated LSD profiles for a given element taking into account the blending of its lines by all other elements. In all cases we also calculated mean profiles from the diagnostic null spectra, applying the same line masks and using the same wavelength regions as in the Stokes $V$ LSD analysis. No evidence of spurious polarisation signatures was found.

The application of LSD led to more than a tenfold increase of the S/N ratio of the Stokes $V$ spectrum, allowing to reach typically $2\times10^4$ per velocity bin (see 5th column in Table~\ref{tbl:obs}). This enabled a secure detection of the circular polarisation signatures for all rotational phases. An example of the LSD Stokes $V$ profile obtained using the full metal line mask is presented in Fig.~\ref{fig:lsd}. This particular observation corresponding to the rotational phase 0.424 yields the highest amplitude circular polarisation signature detected in our data. Interestingly, the Stokes $V$ amplitude reaches maximum for the rotational phase interval 0.3--0.5, when the mean longitudinal magnetic field changes sign. During this part of the rotation cycle the Stokes $V$ profile has a predominantly symmetric shape, indicating the presence of at least two regions of different field polarity on the stellar surface. Such symmetric Stokes $V$ profiles cannot be assessed by the spectropolarimetric techniques limited to \bz\ measurements.

\subsection{Zeeman Doppler inversions with realistic interpretation of LSD profiles}

Numerous previous studies demonstrated the least-squares deconvolution to be a remarkably powerful method for detecting weak stellar magnetic fields and for precise measurements of the mean longitudinal field and, potentially, other field moments. However, the usage of LSD profiles for recovering the physical information about stellar atmosphere or geometrical information about the surface distributions of temperature, chemical abundance, and magnetic field is more controversial. All types of modelling of LSD profiles attempted so far relied on a number of simplifying, often severe, assumptions about the nature of the mean profiles and their relation to the mean line parameters inferred from the LSD line mask.

\begin{figure*}[!th]
\centering
\resizebox{!}{8cm}{\rotatebox{90}{\includegraphics*{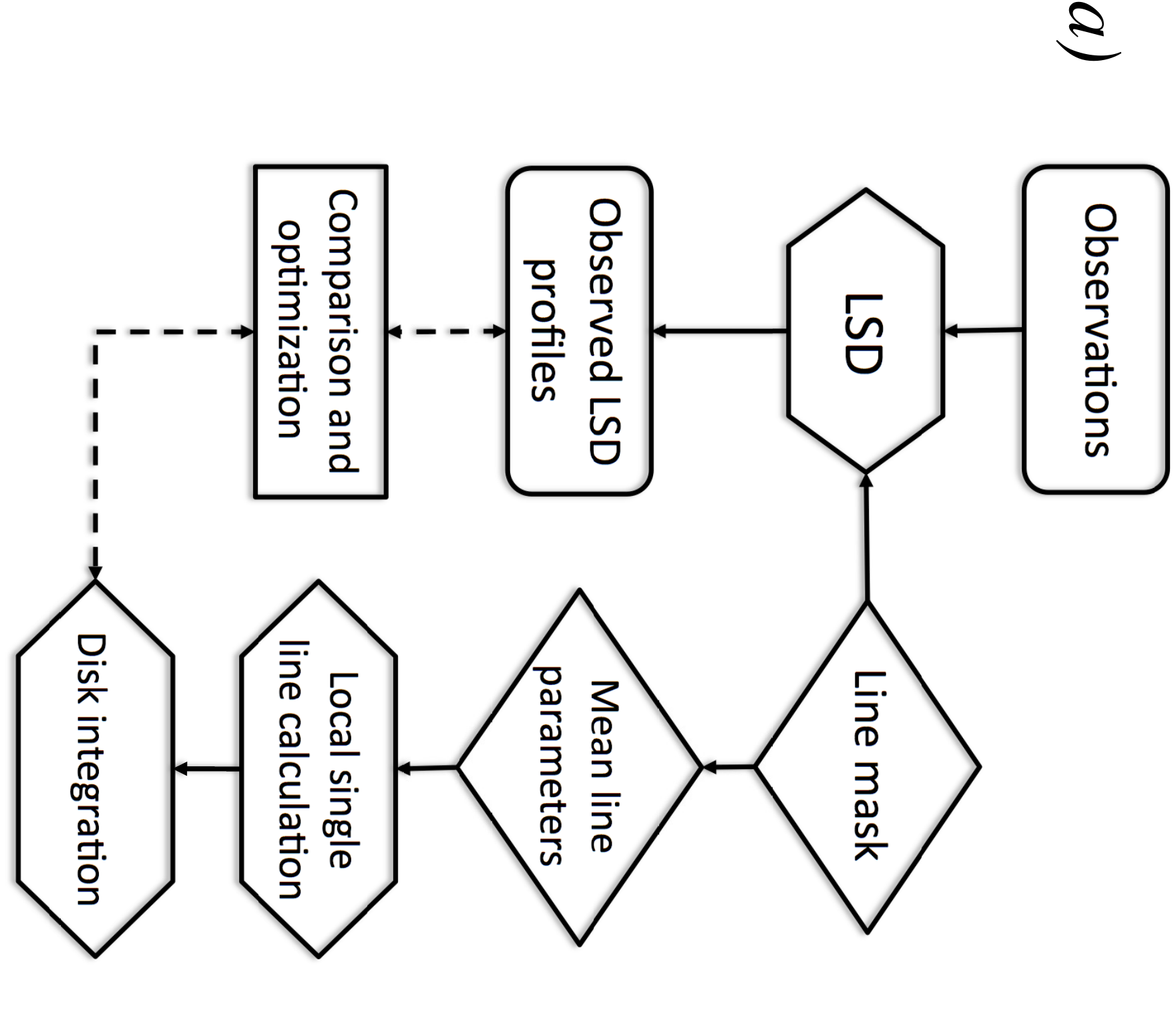}}} \hspace*{0.5cm}
\resizebox{!}{7.9cm}{\rotatebox{90}{\includegraphics*{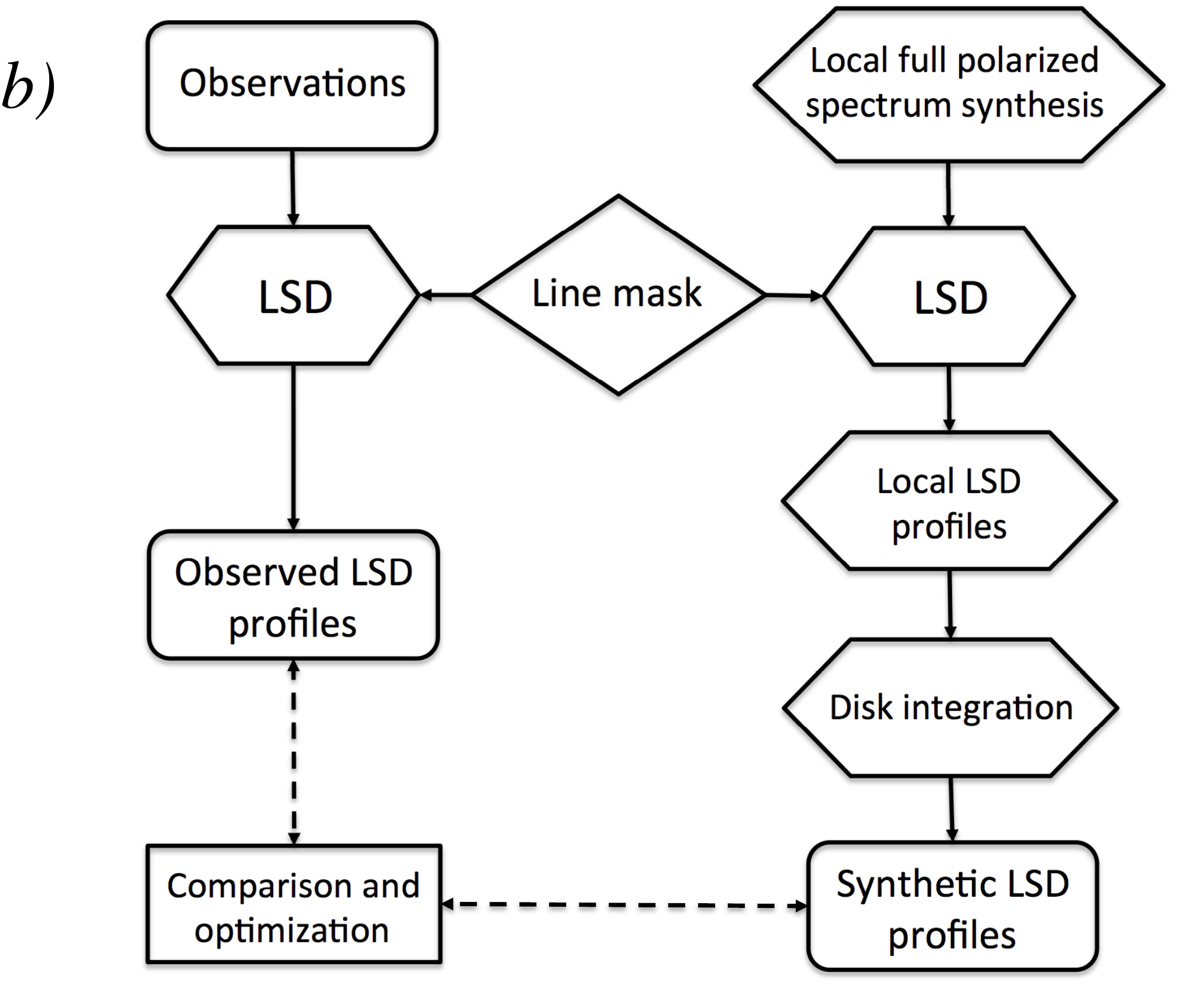}}}
\caption{Two methods of interpreting LSD profiles in ZDI: {\bf a)} traditional single-line approximation,
{\bf b)} new realistic polarized radiative transfer approach developed in this study.}
\label{fig:scheme}
\end{figure*}

In practice, interpretation of LSD profiles in the context of DI or ZDI studies proceeds as follows (see Fig.~\ref{fig:scheme}a). First, one adopts appropriate stellar atmospheric parameters and calculates intensities of spectral features contributing to the LSD line mask. Then LSD profiles are derived for this mask and for given observational data. This step also provides weighted mean values of different atomic parameters corresponding to the LSD mask. On the next step one assumes, in one way or another, that LSD profiles behave similar to a single spectral feature with the previously found mean parameters. In particular, the response to the magnetic field is assumed to be that of a Zeeman triplet with the mean effective Land\'e factor. Using this approximation, local line profiles are computed. Disk integration is performed and a comparison between the observed and computed LSD profiles is carried out.

Previous studies differ in the details of theoretical calculations of the local LSD profiles. Most ZDI investigations of cool stars \citep[e.g.][]{petit:2004a,marsden:2011} assume that the local Stokes $I$ profile is represented by a temperature-independent Gaussian function and the local Stokes $V$ profile is given by the derivative of the Gaussian, according to the weak field approximation. It is not entirely clear how the width and intensity of this Gaussian profile are chosen, especially when the observed LSD profiles are strongly distorted by cool spot signatures. A handful of studies \citep{donati:2006b,morin:2008} used a more sophisticated Unno-Rachkovsky analytical solution of the polarised radiative transfer (PRT) equation for calculating local LSD profiles. This method does not rely on magnetic field being weak, but, again, requires a number of input line parameters that cannot be determined uniquely. Finally, \citet{folsom:2008} and \citet{kochukhov:2013} attempted to approximate the local LSD profiles with a calculation based on a realistic stellar model atmosphere and numerical PRT modelling. In that case it is assumed that the local LSD profile behaves as a single line of the dominant ion (typically \ion{Fe}{i} or \ion{Fe}{ii}) with some parameters (wavelength, excitation potential, Land\'e factor) determined from the LSD mask and other parameters (oscillator strength, damping) adjusted to match the LSD profile shape obtained from a full theoretical spectrum synthesis calculation.

Notwithstanding different degrees of sophistication of various LSD profile modelling approaches used so far, \textit{the single-line approximation represents the most significant limitation of the LSD methodology in the context of DI and ZDI}. Given a wide range of spectral line responses to temperature, abundance, and magnetic field, it is unclear if any single line can match an average profile constructed from hundreds or thousands absorption features. To this end, \citet{kochukhov:2010a} showed that, even with the most detailed PRT interpretation, the validity range of the single-line LSD methodology is very limited. The LSD Stokes $I$ and $V$ profiles can be reasonably matched by an adjusted synthetic line only up to the field strength of $\approx$\,2~kG. The Stokes $Q$ and $U$ profiles cannot be matched at any field strength. Furthermore, the response of the Stokes $I$ LSD profile to chemical abundance variation differs drastically from that of any single line, making an accurate chemical analysis or a realistic abundance DI mapping impossible with LSD profiles. A similar problem affects reconstruction of temperature maps from the observed LSD profiles of cool active stars.

\subsubsection{Theoretical local LSD profiles}

\begin{figure}[!th]
\centering
\figps{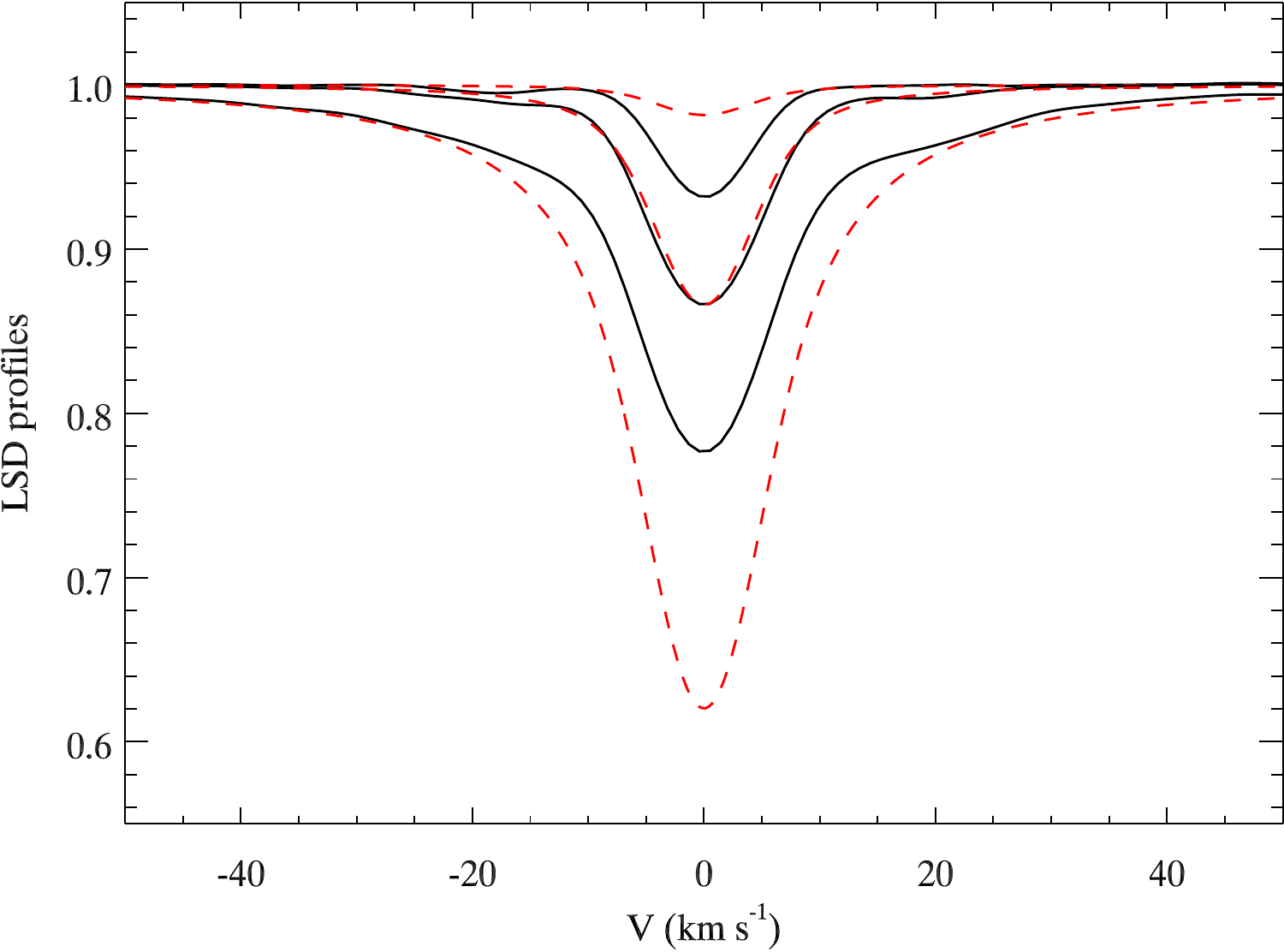}
\caption{Comparison of the synthetic Si LSD profiles (solid lines) with the spectrum synthesis assuming mean line parameters (dashed lines). These calculations show disk-centre local profiles for zero magnetic field and silicon abundances of [Si]\,=\,$-1.0$, 0.0, and +1.0 relative to the Sun. The oscillator strength and Stark damping parameter of the mean line were adjusted to fit the LSD profile corresponding to the solar Si abundance.}
\label{fig:loc}
\end{figure}

In this study we have developed a new method of interpreting LSD profiles that overcomes the main limitations described above. We abandon the single-line LSD profile interpretation in favour of deriving the local LSD profiles from the full spectrum synthesis calculations. This approach, recently applied by \citet{kochukhov:2013b} for forward Stokes $V$ profile calculations, is schematically illustrated in Fig.~\ref{fig:scheme}b. We start by generating a library of theoretical four Stokes parameter spectra, including all relevant absorption lines and covering the entire observed wavelength range, for a grid of abundances/temperatures, magnetic field strengths, orientations of the field vector with respect to the line of sight, and the limb angles. These spectra are convolved with a Gaussian instrumental profile.
Then, LSD profiles are calculated for each of these theoretical spectra using the same line mask as applied to the observed spectra. The relative wavelength dependent weights are taken from a representative observation. Simultaneously, we also determine the mean continuum intensity, which for the case of unpolarised continuum retains only the abundance/temperature and limb angle dependence. The local LSD profiles and continuum intensities for arbitrary parameters are then determined by a linear interpolation within the resulting theoretical LSD profile data base. After interpolation the Stokes $Q$ and $U$ profiles are transformed (rotated) according to the local orientation of the field vector in the plane perpendicular to the line of sight. 
On the final step we perform the disk integration and compare with observations.

The new LSD profile interpretation methodology attributes no particular meaning to the mean profiles. In particular, we do not assume that they behave as a single line, do not require them to be smooth and have certain symmetry properties, and do not impose any limits on the field strength. LSD is merely a method to compress information and numerically compare observations and theoretical calculations. For reasons of numerical tractability the interpretation of LSD Stokes profiles suggested here assumes that it is possible to interchange the operations of disk integration and LSD. This approximation is exact given the linear nature of the disk integration. Its validity has been verified numerically with the help of forward calculations for different star spot scenarios (magnetic, temperature, abundance inhomogeneities).

In the present study of \vir\ we used the {\sc LLmodels} code \citep{shulyak:2004} to generate model atmospheres for two independent grids of Si and Fe abundances. For both elements calculations covered a range from $-2.0$ to $+2.5$ relative to the solar abundance ($\log N_{\rm Si}/N_{\rm tot}=-4.53$ and $\log N_{\rm Fe}/N_{\rm tot}=-4.54$, \citealt{asplund:2009}) with a step of 0.25~dex. Abundances of other elements were either set to the mean values of the DI maps derived by \citet{kuschnig:1999} or assumed to be solar. Theoretical spectra were calculated with the help of the {\sc SynMast} PRT code \citep{kochukhov:2007d,kochukhov:2010a} using a line list containing 2185 lines. This includes all spectral features deeper than 1\% retrieved from the VALD data base for the region of interest. Stokes profiles were evaluated for 17 values of the magnetic field strength between 0 and 7~kG using a step of 250~G for the 0--1~kG interval and a twice larger step for $B>1$~kG. At each field strength value and for each atmospheric model, calculations were performed for a $15\times15$ grid of limb angles and magnetic vector orientations with respect to the line of sight. Theoretical LSD profiles were evaluated for 72675 resulting model Stokes spectra using the same multi-component LSD approach as applied to the observational data.

An example of the synthetic LSD Stokes $I$ profiles for three different Si abundances is shown in Fig.~\ref{fig:loc}. We illustrate an attempt to reproduce these profiles with the single-line spectrum synthesis calculations adjusted to match the intermediate (solar-abundance) Si LSD profile. Even after such an adjustment the single-line spectrum synthesis does not reproduce all details of the reference theoretical LSD profile. Moreover, when the Si abundance is varied by $\pm1$~dex, the single-line calculation fails to correctly predict the LSD profile intensity owing to its much steeper abundance dependence than shown by the theoretical LSD profiles.

\subsubsection{Magnetic and abundance mapping}

The procedure of reconstructing the magnetic field topology and chemical spot maps adopted in our study generally follows that of the {\sc Invers13} code \citep{kochukhov:2012,kochukhov:2013}, with the exception that the local Stokes spectra are obtained by a linear interpolation within the LSD profile table computed as described above rather than calculated on the fly. The stellar surface is divided into 1876 zones with roughly equal areas \citep[see][]{piskunov:2002a}. A chemical map is specified by the element abundance values in each of these surface zones. On the other hand, the maps of radial, meridional, and azimuthal magnetic field components are specified with the help of a general spherical harmonic expansion \citep{donati:2006b}
\beq
B_{\rm r} (\theta, \phi) = -\sum_{\ell=1}^{\ell_{\rm max}} \sum_{m=-\ell}^{\ell} \alpha_{\ell,m} Y_{\ell,m} (\theta, \phi),
\eeq
\beq
B_{\rm m} (\theta, \phi) = -\sum_{\ell=1}^{\ell_{\rm max}} \sum_{m=-\ell}^{\ell} \left[ \beta_{\ell,m} Z_{\ell,m} (\theta, \phi) + \gamma_{\ell,m} X_{\ell,m} (\theta,\phi) \right],
\eeq
\beq
B_{\rm a} (\theta, \phi) = -\sum_{\ell=1}^{\ell_{\rm max}} \sum_{m=-\ell}^{\ell}\left[ \beta_{\ell,m} X_{\ell,m} (\theta, \phi) - \gamma_{\ell,m} Z_{\ell,m} (\theta,\phi) \right],
\eeq
where we use the real spherical harmonic functions to describe the mode with degree $\ell$ and order $m$
\beq
Y_{\ell,m} (\theta, \phi) = -C_{\ell,m} P_{\ell,|m|} (\cos\theta) K_m (\phi),
\eeq
\beq
Z_{\ell,m} (\theta, \phi) = \frac{C_{\ell,m}}{\ell+1} \frac{\partial P_{\ell,|m|} (\cos\theta)}{\partial \theta} K_m (\phi),
\eeq
\beq
X_{\ell,m} (\theta, \phi) = -\frac{C_{\ell,m}}{\ell+1} \frac{P_{\ell,|m|} (\cos\theta)}{\sin \theta} m K_{-m} (\phi),
\eeq
with
\beq
C_{\ell,m} = \sqrt{\frac{2\ell+1}{4\pi}\frac{(\ell-|m|)!}{(\ell+|m|)!}}
\eeq
and
\beq
K_m (\phi) = \left\{  
\begin{array}{ll}
\cos(|m|\phi), & m \ge 0 \\
\sin(|m|\phi), & m < 0 \\
\end{array}\right..
\eeq
Here $\theta$ and $\phi$ are the colatitude and longitude at the stellar surface and $P_{\ell,m} (\theta)$ is the associated Legendre polynomial. Its derivative with respect to $\theta$, required to evaluate $Z_{\ell,m} (\theta,\phi)$, can be computed with
\begingroup
\renewcommand*{\arraystretch}{2.5}
\beq
\begin{array}{rl}
\dfrac{\partial P_{\ell,|m|} (\cos{\theta})}{\partial \theta} = & \dfrac{\ell - |m| +1}{\sin\theta} P_{\ell+1,|m|} (\cos{\theta}) \\
& - \dfrac{\ell+1}{\sin\theta} \cos\theta P_{\ell,|m|} (\cos{\theta}).
\end{array}
\eeq
\endgroup
The free parameters of this harmonic expansion are coefficients $\alpha_{\ell,m}$, $\beta_{\ell,m}$, and $\gamma_{\ell,m}$. They characterise contributions of the radial poloidal, horizontal poloidal, and  horizontal toroidal magnetic field components, respectively. The expansion is carried out up to a sufficiently large $\ell_{\rm max}$ as to fit the details of the observed Stokes profiles. In the present study of \vir\ we used $\ell_{\rm max}=10$, yielding 360 independent magnetic variables.

An iterative reconstruction of the abundance and magnetic maps, aimed to fit the observed Stokes $I$ and $V$ profiles, is carried out simultaneously and self-consistently, using the modified Levenberg-Marquardt optimisation algorithm \citep{piskunov:2002a}. For the regularisation, which is a necessary ingredient of any Doppler imaging code, we use the Tikhonov functional \citep[e.g.][]{tikhonov:1977} for the abundance map and a penalty function $\sum_{\ell,m} \ell^2 (\alpha_{\ell,m}^2 + \beta_{\ell,m}^2 +\gamma_{\ell,m}^2)$ for the magnetic field, with corresponding individual regularisation parameters. This means that we constrain the inversion by requiring the final abundance map to be as smooth as allowed by observations and the magnetic map to contain the least possible contribution of the higher-order harmonic modes. Alternatively, the magnetic penalty function can be substituted with a conservative field parameterisation. For example, the classical oblique dipolar field is realised by setting $\alpha_{\ell,m}= \beta_{\ell,m}$ and $\gamma_{\ell,m}=0$ for $\ell_{\rm max}=1$.

In comparison to the direct reconstruction of magnetic field practiced in our previous ZDI studies of Ap stars \citep{kochukhov:2004d,kochukhov:2010}, the usage of spherical harmonic expansion has the advantage of yielding the null magnetic flux through the stellar surface and hence automatically satisfying the Maxwell's equations. In addition, since individual terms of the expansion are orthogonal, one can conveniently quantify contributions of different spatial scales to the stellar magnetic field topology by comparing the relative energies ($B^2$ integrated over the stellar surface) of harmonic components. The disadvantage of the harmonic expansion and the associated penalty function is that this method is difficult to apply to stars showing fine-structured bipolar field geometries. However, this problem concerns only certain types of active cool stars rather than early-type magnetic stars, such as the one studied here.

\section{Results}
\label{results}

\subsection{Stellar parameters}
\label{params}

Determination of the atmospheric parameters of magnetic CP stars is notoriously difficult due to non-solar chemical abundance patterns and variability. Using standard photometric \teff\ calibrations trained on normal stars may lead to significant systematic errors, especially for the Si-type CP stars such as \vir\ \citep{netopil:2008}. In this situation it is preferable to rely on more detailed model atmosphere analyses, which attempt to account for chemical peculiarity, at least to some extent. To this end, the most relevant previous studies of \vir\ are those by \citet{kuschnig:1999}, \citet{shulyak:2004}, and \citet{lipski:2008}. The first of these publications suggested \teff\,=\,13000~K and \lgg\,=\,4.0 based on spectroscopy. This set of parameters was also used by \citet{krticka:2012} in the modelling of the optical and ultraviolet spectral energy distribution (SED) of \vir. On the other hand, \citet{shulyak:2004} and \citet{lipski:2008} directly fitted the observed SED, finding \teff\,=\,12750~K and \lgg\,=\,4.0--4.5. Taking into account these results, we adopted \teff\,=\,$12750\pm250$~K in the present paper.

\begin{figure}[!th]
\centering
\figps{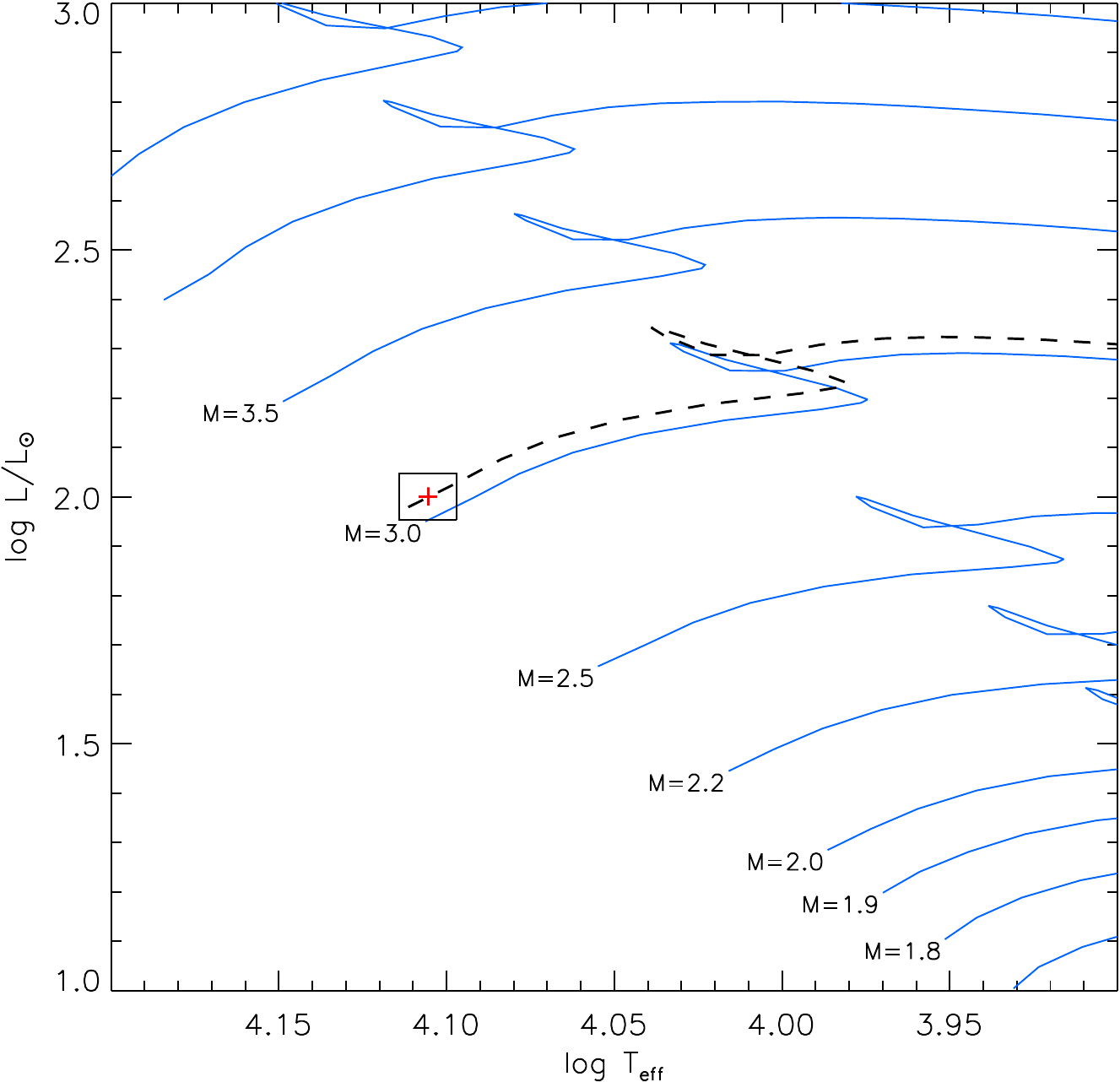}
\caption{Comparison of the H-R diagram position of \vir\ (box) with the Padova theoretical stellar evolutionary tracks (solid lines). The dashed line shows an evolutionary track interpolated for $M=3.06M_\odot$. 
}
\label{fig:tracks}
\end{figure}

Combining \teff\ with the observed $V$ magnitude, the trigonometric parallax $12.63\pm0.21$~mas \citep{van-leeuwen:2007}, and an empirical bolometric correction $BC=-0.79\pm0.1$ \citep{lipski:2008}, we found the stellar luminosity $L=100\pm11$\,$L_\odot$ and the radius $R=2.06\pm0.14$\,$R_\odot$. Then, a comparison with the Padova theoretical stellar evolutionary tracks \citep{bertelli:2009} 
allows to place \vir\ in the H-R diagram. As illustrated in Fig.~\ref{fig:tracks}, the parameters of our target correspond to the stellar mass $M=3.06\pm0.06$\,$M_\odot$. The surface gravity is \lgg\,=\,$4.30\pm0.06$ and the stellar age is constrained to be $\le$\,100~Myr. The stellar radius obtained here is consistent with $R=2.3\pm0.1$\,$R_\odot$ \citep{kochukhov:2006} and $R=2.2\pm0.2$\,$R_\odot$ \citep{north:1998a} derived from photometry and evolutionary tracks. A similar radius of $R=1.9\pm0.1$\,$R_\odot$ was estimated by \citet{krticka:2012} from the observed UV flux. \citet{lipski:2008} also reported $R=2.2$\,$R_\odot$ from the optical SED.

\begin{table}[!t]
\caption{Parameters of \vir.
\label{tbl:params}}
\begin{tabular}{ll}
\hline\hline
Parameter & Value \\
\hline
$T_{\rm eff}$ & $12750\pm250$ K \\
$\log g$ & $4.30\pm0.06$ \\
$L/L_\odot$ & $100\pm11$ \\
$R/R_\odot$ & $2.06\pm0.14$ \\
$M/M_\odot$ & $3.06\pm0.06$ \\
\vs\ & $145\pm3$ \kms\ \\
$P_{\rm rot}$ & 0.5207130~d \\
$i$ & $46.5\pm4.1\degr$ \\
\hline
\end{tabular}
\end{table}

\begin{figure}[!th]
\centering
\figps{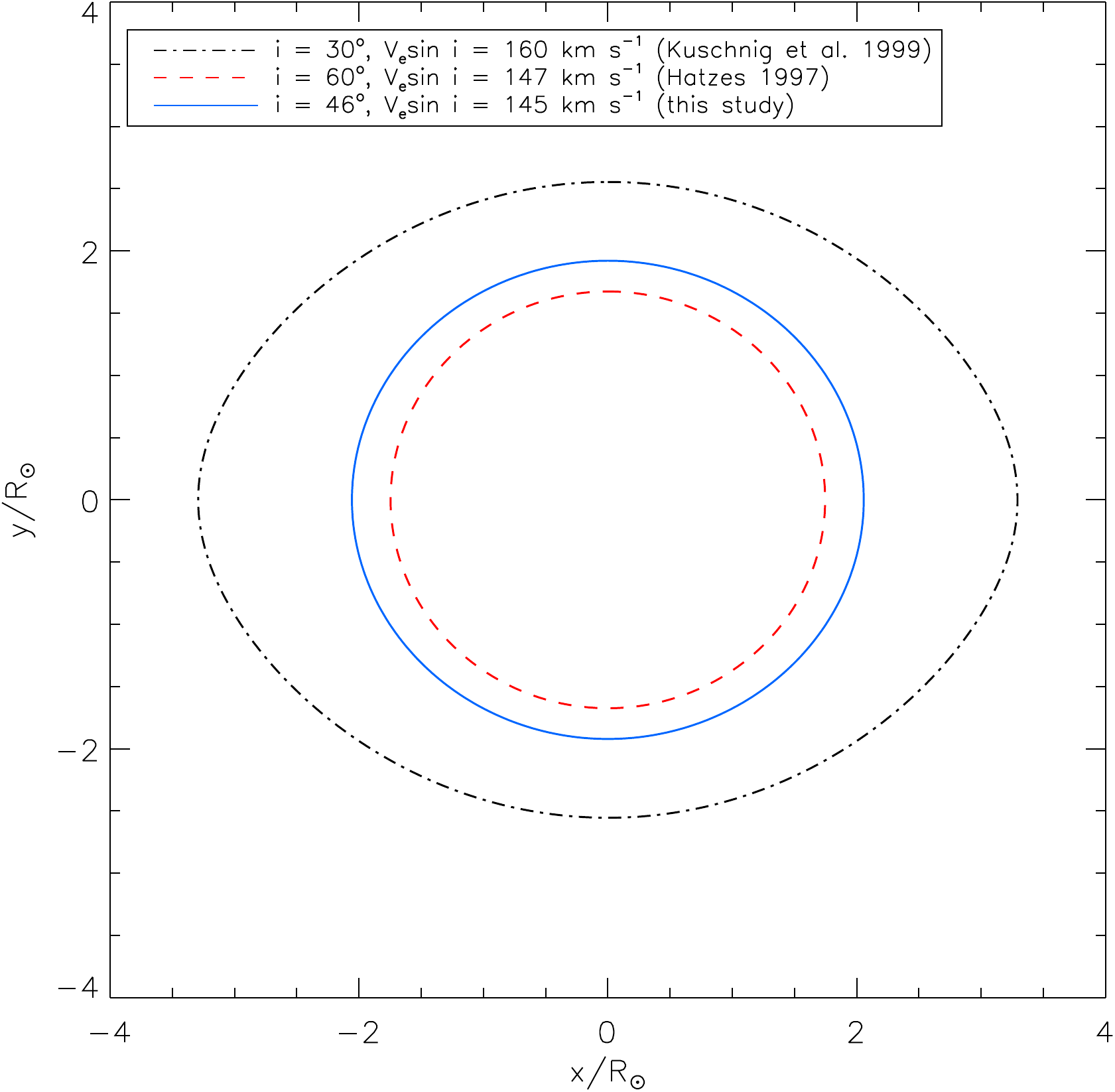}
\caption{Effect of fast rotation on the surface shape of \vir. The stellar shape is shown here for \teff\,=\,12750~K, $M=3.06M_\odot$, $P_{\rm rot}=0.5207130$~d and different combinations of $i$ and \vs\ adopted in the Doppler imaging studies of this star. The stellar rotational axis corresponds to a line at $x=0$.}
\label{fig:shape}
\end{figure}

Taking into account the projected rotational velocity \vs\,=\,$145\pm3$~\kms\ obtained from the DI inversions in this paper, we can refine the inclination angle to $i=46.5\pm4.1\degr$ using the oblique rotator relation. This is close to $i=43\pm7\degr$ estimated by \citet{trigilio:2000} and compatible with \vs\,=\,$147\pm2$~\kms, $i=60\degr$ \citep{hatzes:1997} and with \vs\,=\,150~\kms, $i=60\degr$ \citep{hiesberger:1995} given a typical uncertainty of 10--20\degr\ for the inclination derived by DI fits. On the other hand, \citet{kuschnig:1999} adopted \vs\,=\,160~\kms\  and $i=30\degr$, yielding a stellar radius of $R=3.3$\,$R_\odot$. This radius is substantially larger than other estimates and is unexpected for an unevolved main sequence star with the parameters of \vir.

Different sets of \vs, $i$, and $R$ discussed here correspond to different equatorial rotational velocities and hence a different impact of rapid rotation on the observed stellar properties. Using standard relations describing the shape of rapidly rotating stars and the corresponding gravity darkening \citep[e.g.][]{collins:1963}, we assessed the implication of different $v_{\rm e}$. Figure~\ref{fig:shape} shows that the set of parameters used by \citet{kuschnig:1999} actually yields a noticeable rotational distortion of the stellar surface shape. In that case the equatorial velocity amounts to 76 per cent of the critical value and the pole-to-equator temperature difference reaches 3700~K. However, these effects were ignored by \citet{kuschnig:1999}, which makes their DI inversions inconsistent with the basic stellar parameters adopted in their study. On the other hand, the smaller $R$ and lower \vs\ adopted by us and by \citet{hatzes:1997} result in a nearly spherical stellar surface, $v_{\rm e}/v_{\rm crit}\le0.37$, and a much lower temperature surface variation (500--800~K), which can be neglected in DI of \vir.

Table~\ref{tbl:params} summarises all parameters of \vir\ assumed or derived in our study. 

\subsection{Longitudinal magnetic field variation}
\label{longit}

The mean line of sight magnetic field, \bz, was obtained from the normalised first moment of the Stokes $V$ LSD profiles, as described by \citet{kochukhov:2010a}. For this analysis we used the LSD profiles computed using all metal lines. These measurements, provided in Table~\ref{tbl:obs}, show a smooth, single-wave rotational variation, with \bz\ changing between about 900~G and $-700$~G. The median error of our mean longitudinal field estimates is 60~G.

\begin{figure}[!th]
\centering
\figps{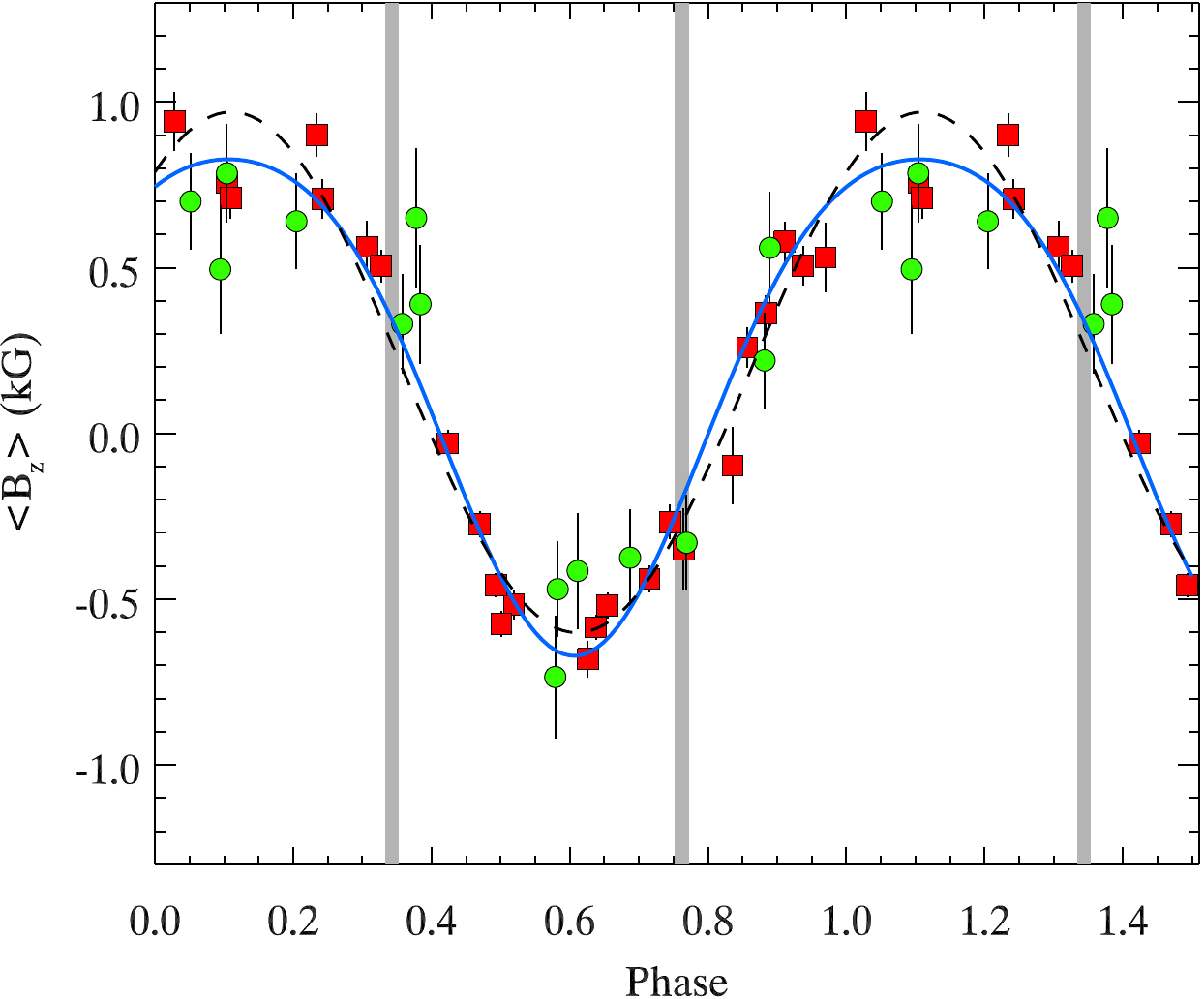}
\caption{Variation of the mean longitudinal magnetic field of \vir. The squares show \bz\ measurements derived in this study from the LSD profiles of all metal lines. The circles show Balmer line photopolarimetric measurements by \citet{borra:1980}. The solid line corresponds to the fit with a dipole plus quadrupole magnetic field model. The dashed line shows predictions of the best-fitting pure dipolar model. The thick vertical grey lines indicate rotational phases of the peak radio frequency emission at 1.4 GHz.}
\label{fig:bz}
\end{figure}

In Fig.~\ref{fig:bz} we compare our \bz\ measurements derived from the high-resolution Narval spectra with the photopolarimetric magnetic observations by \citet{borra:1980} obtained more than 30 years ago from the hydrogen Balmer line wings. Despite using different diagnostic techniques and different spectral features, the two data sets show a reasonable agreement when phased together using the ephemeris with variable rotational period \citep{mikulasek:2011}. At the same time, our metal line \bz\ measurements are on average 2.5 times more precise and indicate a slightly higher amplitude of the longitudinal field variation.

\begin{figure}[!th]
\centering
\fifps{8cm}{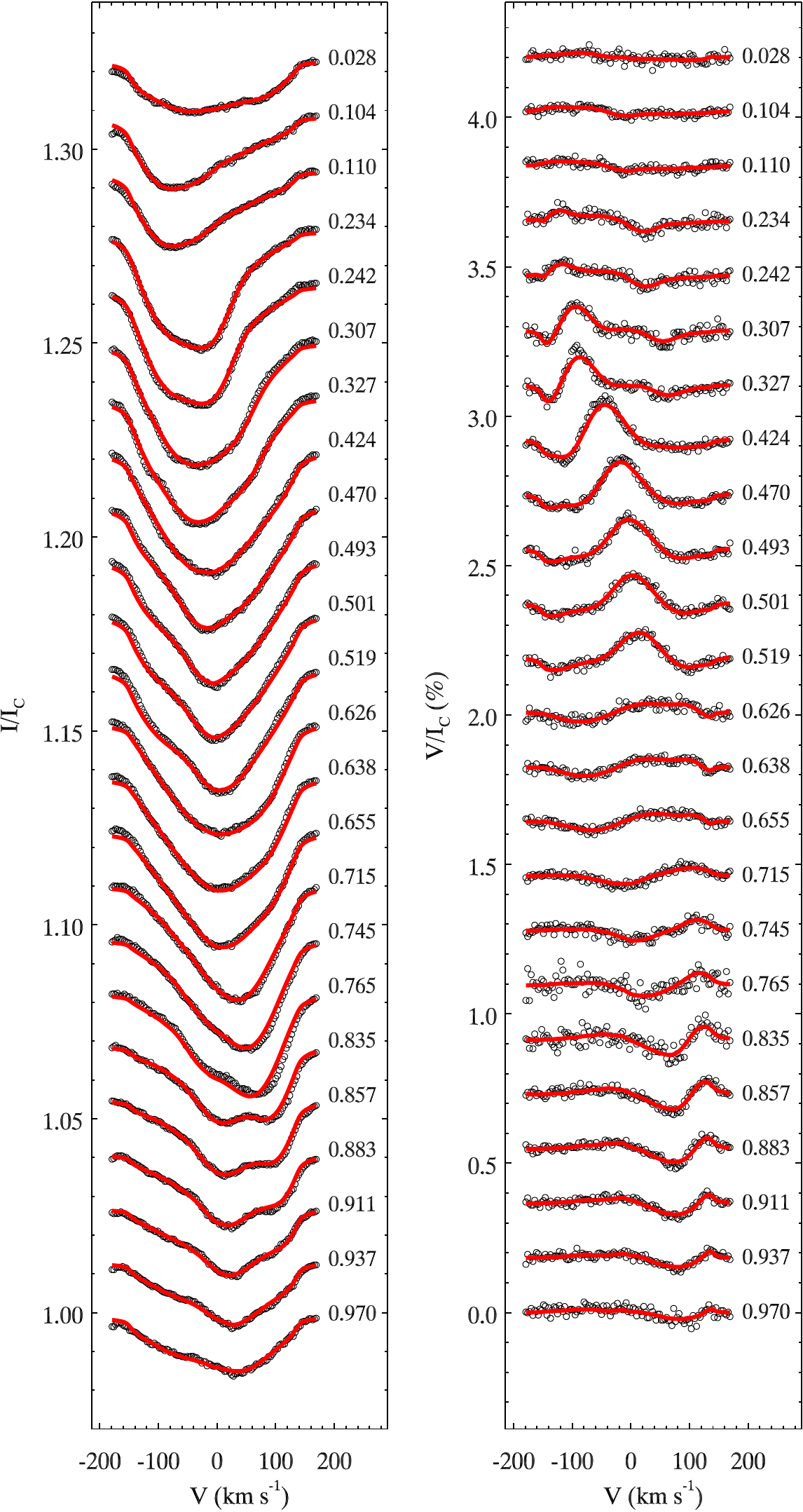}
\caption{Comparison of the observed Si LSD Stokes $I$ (left) and $V$ (right) profiles of \vir\ with the fit achieved by the magnetic inversion code. Observations are shown with symbols. Calculations for the final surface distribution of Si spots and the best-fit magnetic field model are shown with the thick red line. 
Spectra corresponding to different rotation phases are offset vertically. Rotation phases are indicated to the right of each spectrum.}
\label{fig:prf_si}
\end{figure}

\begin{figure}[!th]
\centering
\fifps{8cm}{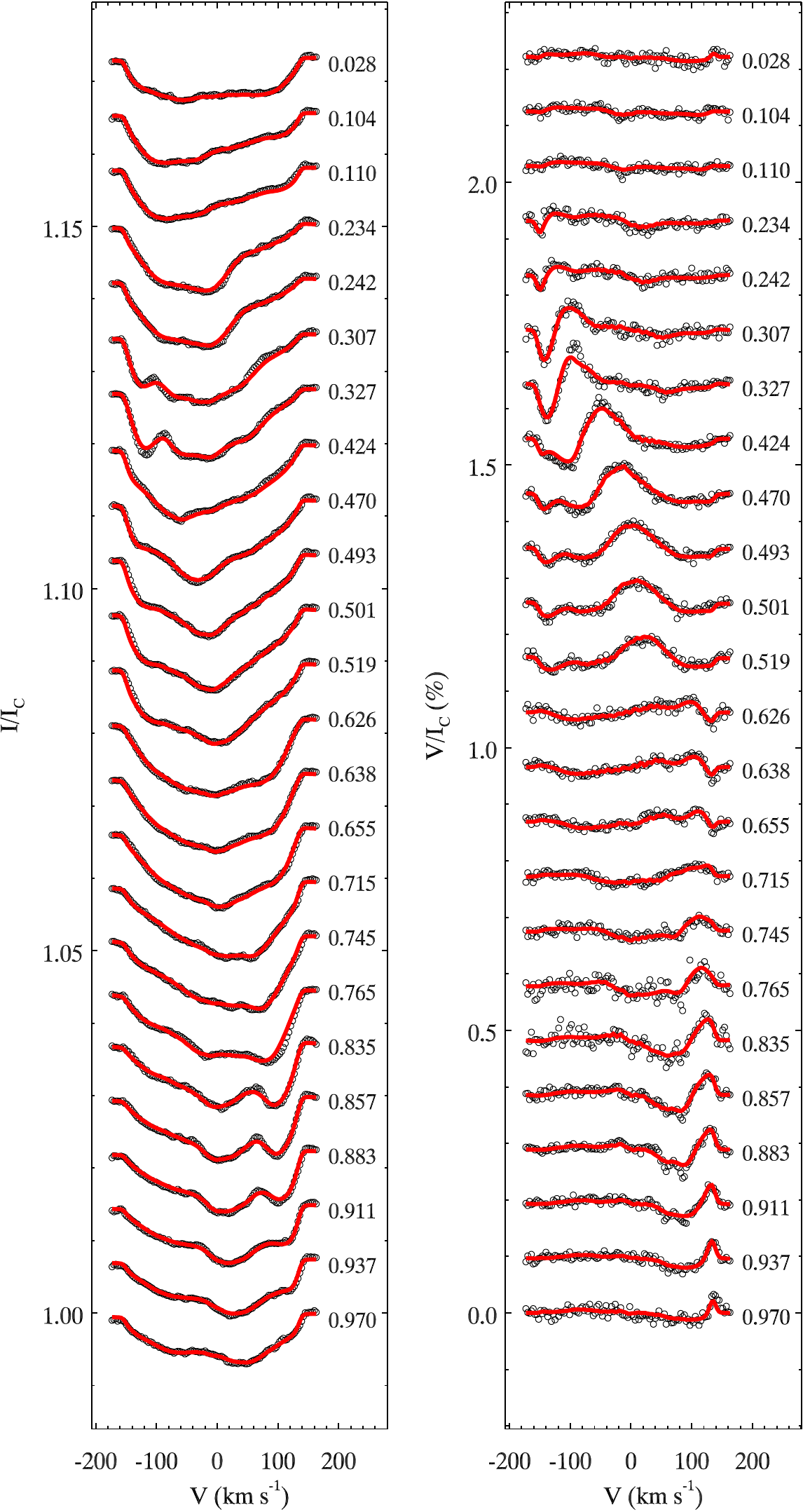}
\caption{Same as Fig.~\ref{fig:prf_si} for the observed and synthetic Fe LSD profiles.}
\label{fig:prf_fe}
\end{figure}

Using the times of radio emission maxima at 1.4~GHz published by \citet{trigilio:2000,trigilio:2011} and \citet{leto:2006}, we can establish that the peaks of radio emission occur at the rotational phases $0.34\pm0.01$ and $0.76\pm0.01$ for the variable-period ephemeris adopted in our study. These phases are not distributed symmetrically with respect to any of the two \bz\ extrema and are offset by $-0.07$ (first peak) and $-0.04$ (second peak) from the phases of the zero longitudinal magnetic field.

The longitudinal field curve of \vir\ exhibits an asymmetry between the two magnetic poles: the negative field extremum is somewhat sharper than the positive field maximum. This indicates a departure from a purely dipolar field configuration: e.g. the presence of an offset dipolar field or a superposition of the dipolar and quadrupolar magnetic components. Adopting the latter field parameterisation and assuming that the quadrupole is axisymmetric and aligned with the dipole \citep[e.g.][]{landstreet:2000}, we obtained $B_{\rm d}=4.0\pm0.2$~kG, $B_{\rm q}=-7.6\pm2.2$~kG, and $\beta=79\pm2$\degr\ for a fixed inclination angle $i=46.5$\degr. The resulting fit to the longitudinal field curve (solid curve in Fig.~\ref{fig:bz}) is generally satisfactory although not perfect, yielding $\chi^2_{\nu}=2.9$. The remaining discrepancies are probably related to the effects of chemical spots and brightness variation across the stellar surface. A pure oblique dipolar model gives $B_{\rm d}=3.8\pm0.2$~kG, $\beta=76\pm2$\degr\ and corresponds to a significantly worse fit ($\chi^2_{\nu}=4.3$, dashed curve in Fig.~\ref{fig:bz}) to the new \bz\ measurements. In any case, the value of $\beta$ agrees well with the results of previous determinations \citep{hatzes:1997,trigilio:2000} using the \bz\ data of \citet{borra:1980}.

Notwithstanding the improvement of the description of \bz\ variation brought by considering quadrupolar terms, the magnetic field geometry inferred from the longitudinal field curve appears to be misleading. As we will show below, this field configuration does not reproduce the observed Stokes $V$ profile behaviour and therefore cannot be accepted as a realistic representation of the magnetic field structure at the surface of \vir.

\subsection{Magnetic field topology and chemical abundance distributions}

\begin{figure*}[!th]
\centering
\firrps{15cm}{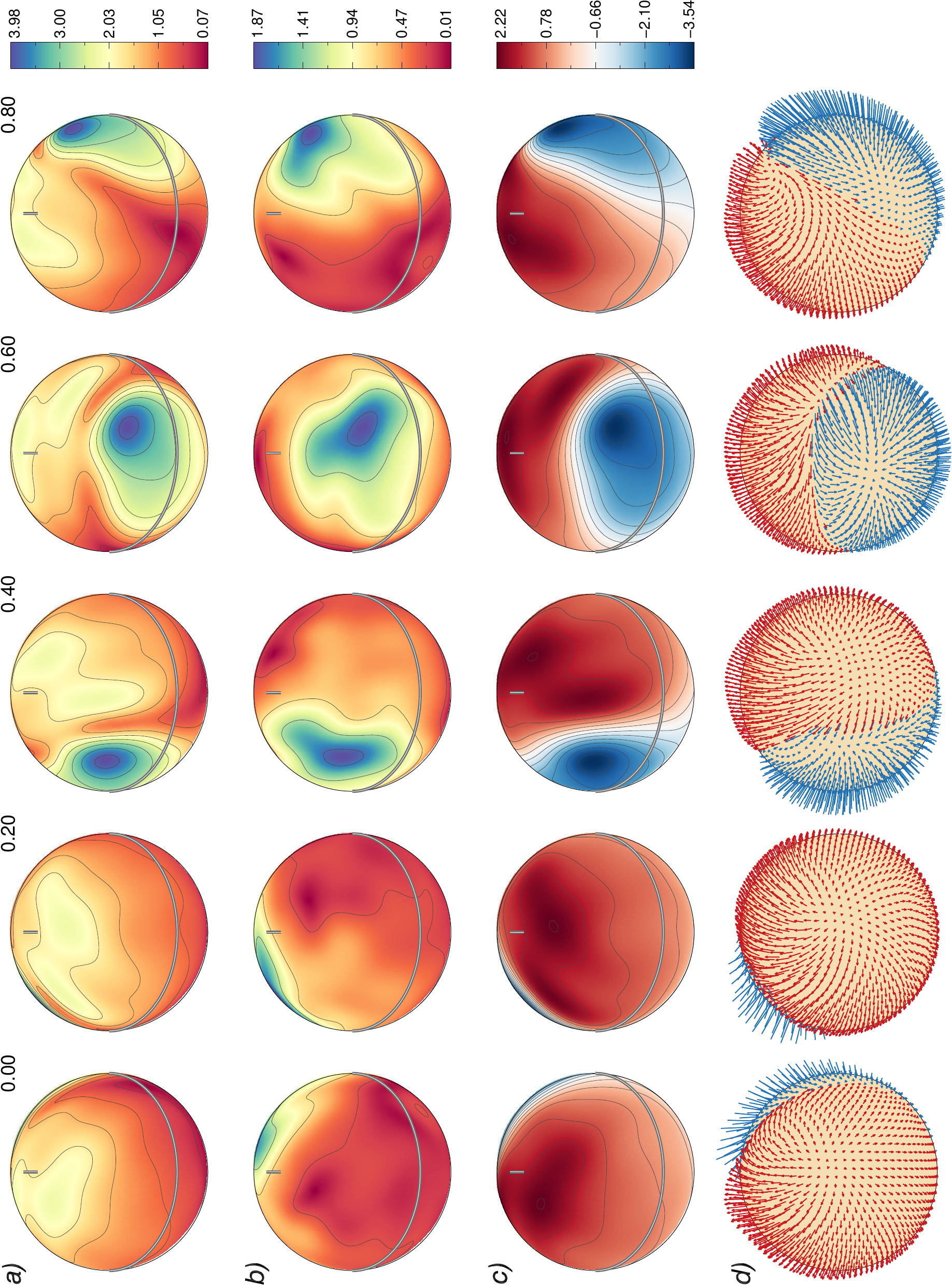}
\caption{Surface magnetic field distribution of \vir\ derived from Si LSD profiles. The star is shown at five rotation phases, indicated above the spherical plots, and the inclination angle of $i=46.5\degr$. The spherical plots show the maps of {\bf a)} field modulus, {\bf b)} horizontal field, {\bf c)} radial field, and {\bf d)} field orientation. The contours over spherical maps are plotted with a 0.5~kG step. The thick line and the vertical bar indicate positions of the rotational equator and the pole, respectively. The colour bars give the field strength in kG. The two different colours in the field orientation map correspond to the field vectors directed outwards (red) and inwards (blue).}
\label{fig:fld_si}
\end{figure*}

ZDI reconstruction of the magnetic field topology and abundance distributions was performed separately for the Si and Fe LSD profiles. In Figs.~\ref{fig:prf_si} and \ref{fig:prf_fe} we compare the observed Stokes $I$ and $V$ spectra with the final fit achieved by the magnetic inversion code. Results of magnetic inversion using the Si LSD profiles are illustrated in Fig.~\ref{fig:fld_si}. This plot shows spherical projections of the field modulus ($\sqrt{B^2_{\rm r} + B^2_{\rm m} + B^2_{\rm a}}$) as well as the horizontal ($\sqrt{B^2_{\rm m} + B^2_{\rm a}}$) and radial ($B_{\rm r}$) field components. The bottom row shows the vector magnetic field map, highlighting the regions of different field polarity.

We find that the magnetic field topology of \vir\ is dipolar-like in the sense that the distribution of the radial field component is dominated by two regions of opposite field polarity. However, other details of the field topology deviate significantly from what is expected for a simple oblique dipolar magnetic field. The maximum local field strength of 4.0~kG is found for the negative radial field region visible at the rotational phase $\approx$\,0.6. On the other hand, the positive longitudinal field extremum corresponds to an extended surface region with a relatively low field strength of 1--2~kG. In other words, \vir\ exhibits a large asymmetry between the negative and positive magnetic poles. The field topology is also clearly non-axisymmetric, with the strongest positive and negative radial field regions separated by about 60\degr\ in longitude. According to our inversion results, it is the presence of both regions on the stellar disk at the rotational phase interval 0.3--0.5 that is responsible for a strong crossover signature in the observed LSD Stokes $V$ profiles of \vir.

\begin{figure*}[!th]
\centering
\firrps{15cm}{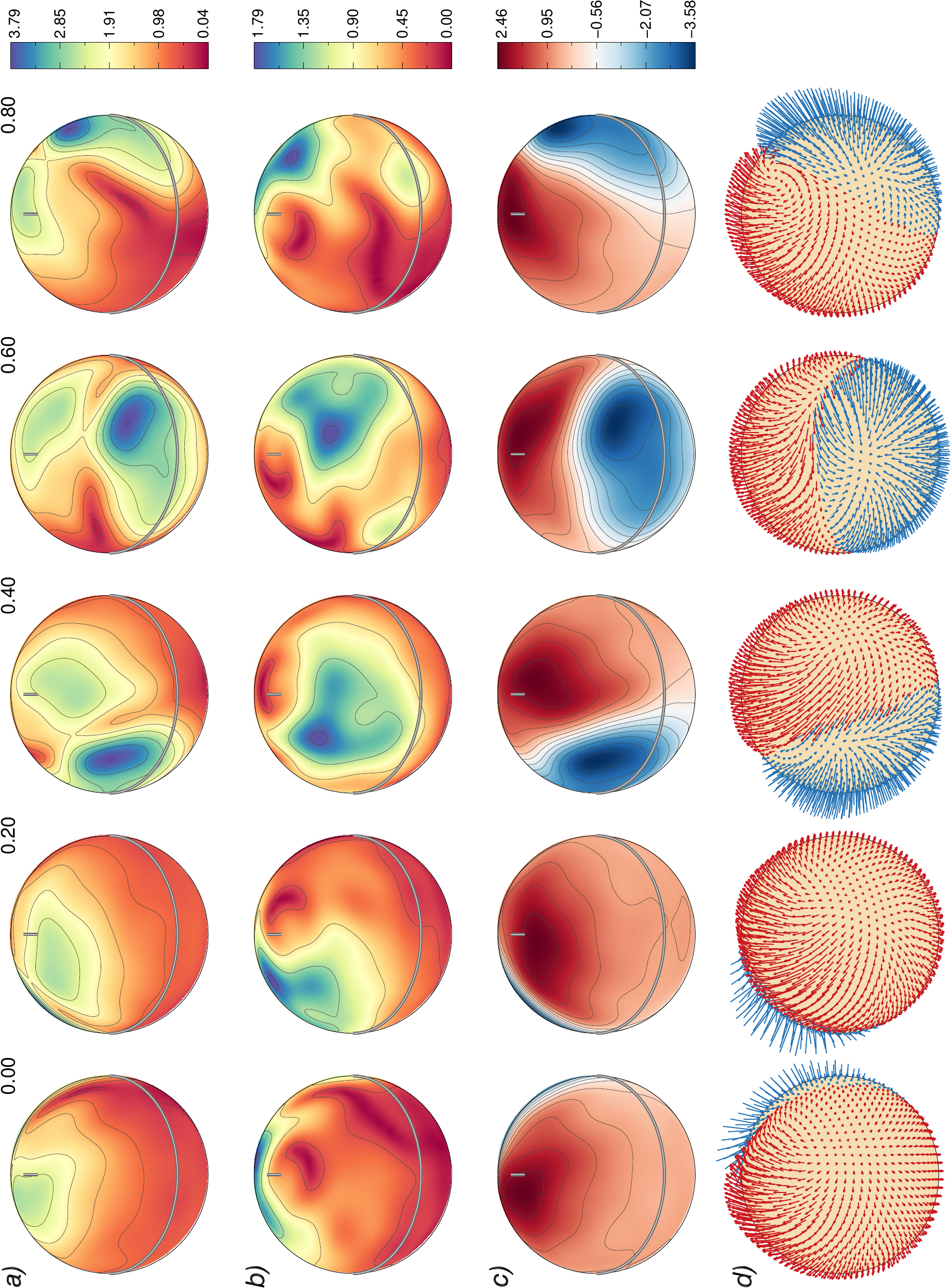}
\caption{Same as Fig.~\ref{fig:fld_si} for the surface magnetic field distribution derived from Fe LSD profiles.}
\label{fig:fld_fe}
\end{figure*}

\begin{figure}[!t]
\centering
{\resizebox{\hsize}{!}{\rotatebox{0}{\includegraphics*{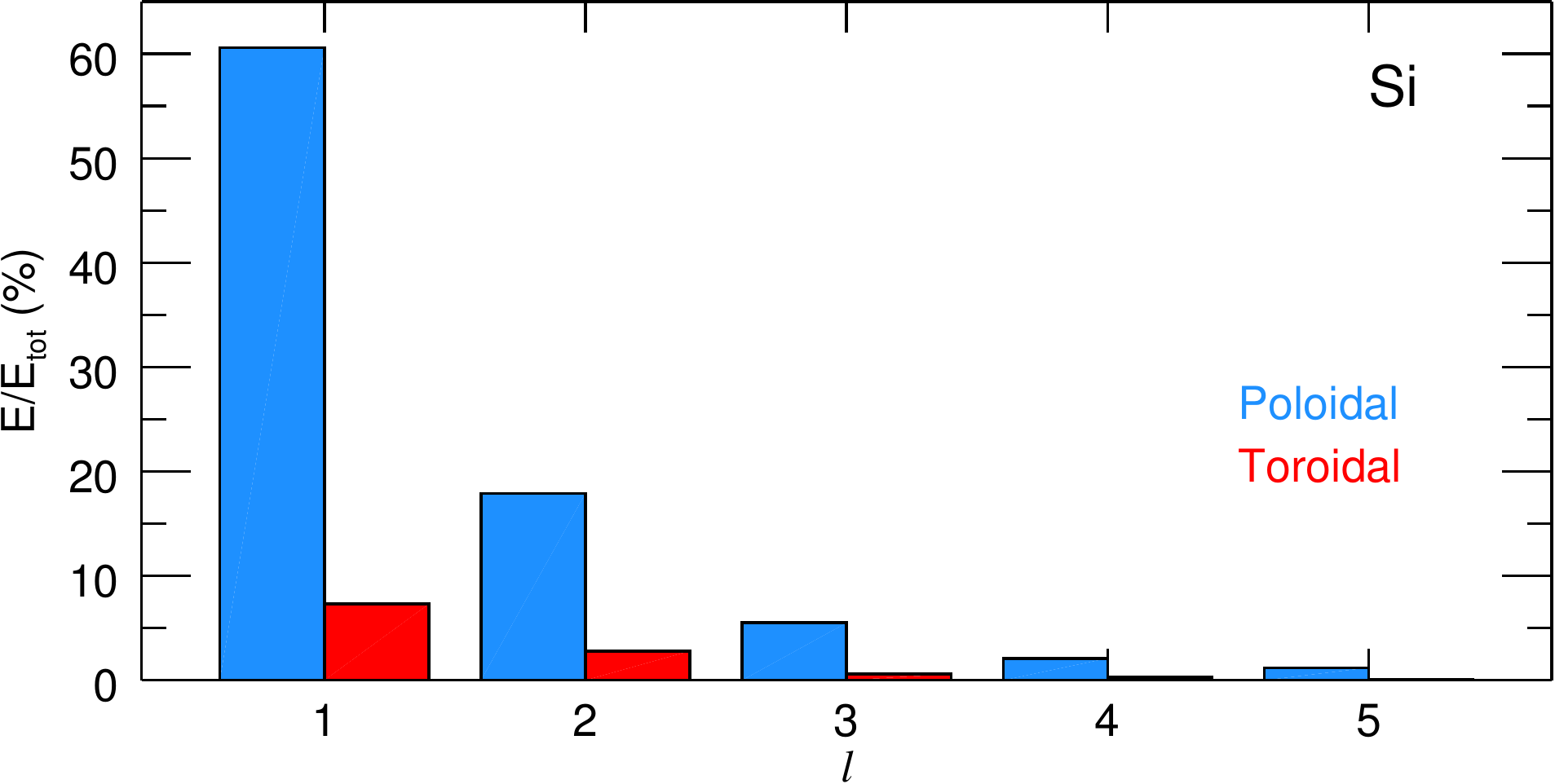}}}}\\\vspace*{0.5cm}
{\resizebox{\hsize}{!}{\rotatebox{0}{\includegraphics*{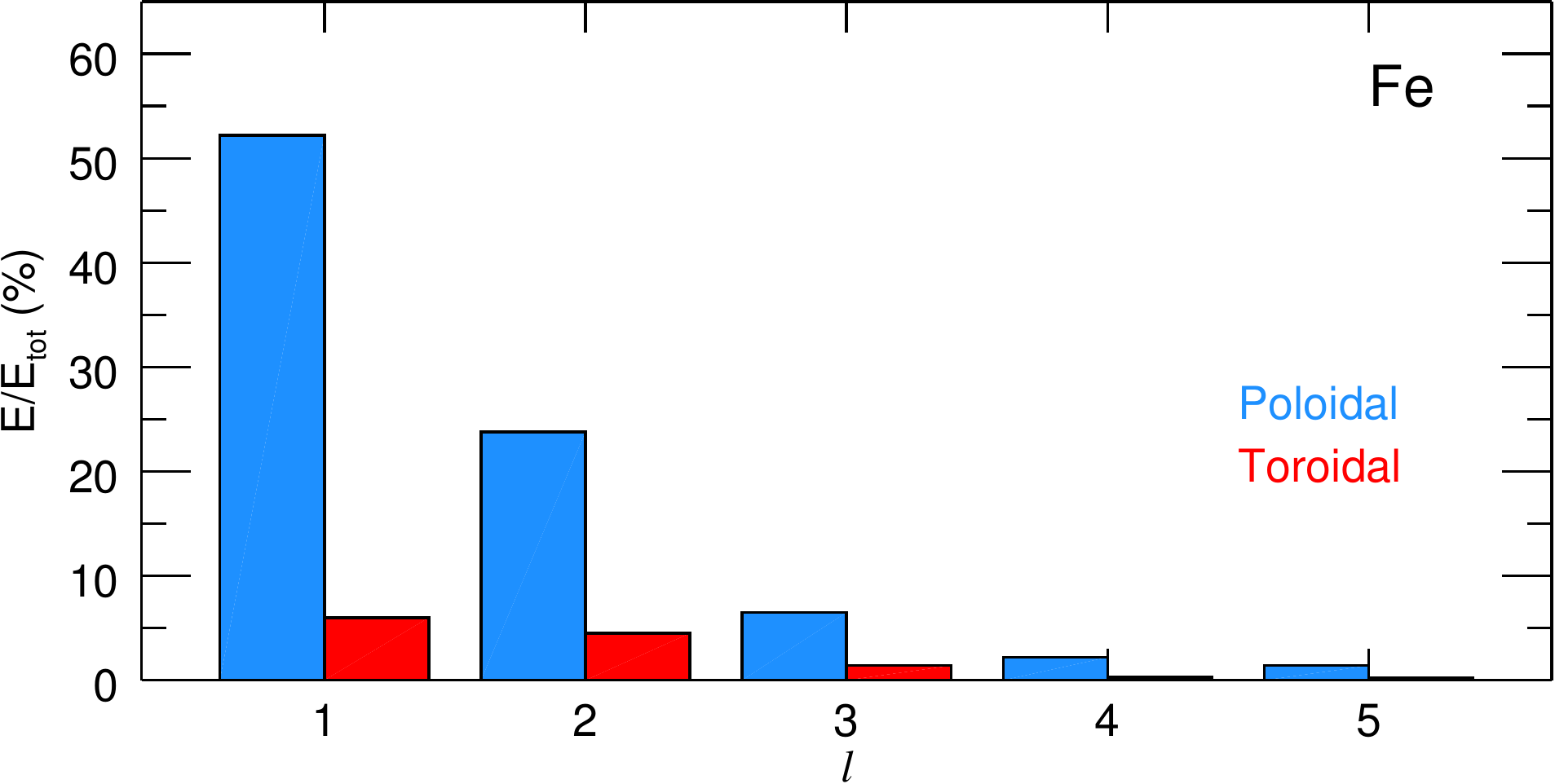}}}}
\caption{Relative energies of the poloidal and toroidal harmonic modes with different angular degrees $\ell$ for the magnetic field topology of \vir\ derived from Si (top) and Fe (bottom) LSD profiles.}
\label{fig:fld_coef}
\end{figure}

The outcome of the magnetic inversions using the Fe Stokes $I$ and $V$ LSD profiles is presented in Fig.~\ref{fig:fld_fe} using the same format as for the Si magnetic maps in Fig.~\ref{fig:fld_si}. The Fe magnetic maps agree closely with the results of Si ZDI inversion. Magnetic maps derived from the mean Fe line exhibits the same field strength distribution asymmetry, the same non-axisymmetric character and similar ranges in the maps of different magnetic field components. We emphasise that analysis of the two sets of LSD profiles was carried out completely independently, starting from zero magnetic field in both cases. The Fe profile inversion results support all aspects of the magnetic field topology found with the ZDI using Si LSD profiles. This confirms robustness of the new ZDI inversion procedure and rules out a possibility of significant random artefacts in the resulting magnetic maps.

As discussed above, ZDI with spherical harmonic parameterisation of the field topology allows one to quantify the contribution of different harmonic modes to the stellar magnetic field topology. In particular, we consider the relative energies obtained by integrating $B^2$ of individual harmonic components over the whole stellar surface. Figure ~\ref{fig:fld_coef} compares the energies of the poloidal and toroidal harmonic modes with $\ell$\,=\,1--5. The poloidal components dominate over the toroidal ones. The dipolar field comprises 52--60\% of the total magnetic field energy. The quadrupolar ($\ell=2$) and octupolar ($\ell=3$) modes are non-negligible, yielding together 23--30\% of the total magnetic energy. The contribution of modes with $\ell\ge4$ is insignificant.

\begin{figure}[!th]
\centering
\fifps{8cm}{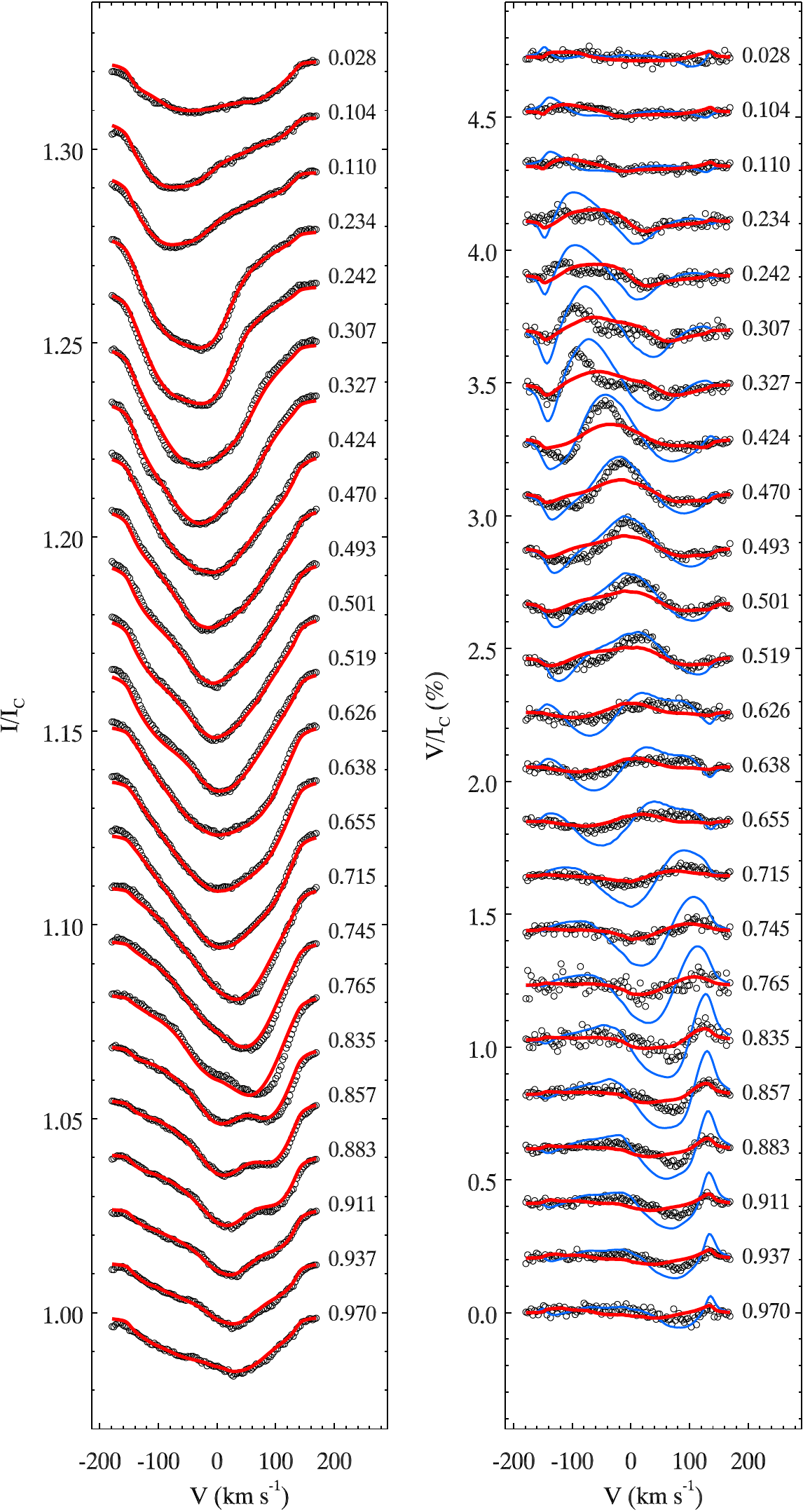}
\caption{Same as Fig.~\ref{fig:prf_si} for the Si LSD profiles modelled assuming a dipolar field topology (thick red line) and for the dipole plus quadrupole field model inferred from the longitudinal field curve (thin blue line).}
\label{fig:prf_si_test}
\end{figure}

The penalty function applied to minimise the unwarranted contribution of the high-order harmonic modes ensures that our ZDI inversion code recovers the simplest possible magnetic field geometry compatible with observations. Nevertheless, to verify the necessity of including complex field components we carried out additional inversions in which we severely limited the maximum angular degree $\ell$ of the harmonic expansion. In particular, we attempted to fit the observed Si LSD profiles with an arbitrary abundance map and a pure dipolar poloidal field topology. As demonstrated by Fig.~\ref{fig:prf_si_test}, this model provides a good description of the Stokes $I$ spectra, but fails to reproduce the circular polarisation profiles of \vir. Theoretical Stokes $V$ profiles cannot match observations in the rotational phase interval 0.3--0.5 when the Stokes $V$ amplitude reaches maximum. The optimal dipolar field parameters ($B_{\rm d}=3.0$~kG and $\beta=57\degr$) suggested by this LSD profile fit are different from the corresponding parameters ($B_{\rm d}=3.8$~kG and $\beta=76\degr$) inferred from the observed \bz\ curve.

Figure~\ref{fig:prf_si_test} also shows that the axisymmetric dipole plus quadrupole field model obtained by fitting the longitudinal field curve is inadequate for describing the high-resolution polarisation observations of \vir. The corresponding theoretical Stokes $V$ profile variation roughly reproduces our observations in the phase interval 0.4--0.5, but also predicts strong and sharp polarisation signatures at phases 0.2--0.4 and 0.6--0.8. No such features are seen in the observed Stokes $V$ spectra, either in individual lines or in LSD profiles.

\begin{figure*}[!th]
\centering
\firrps{15cm}{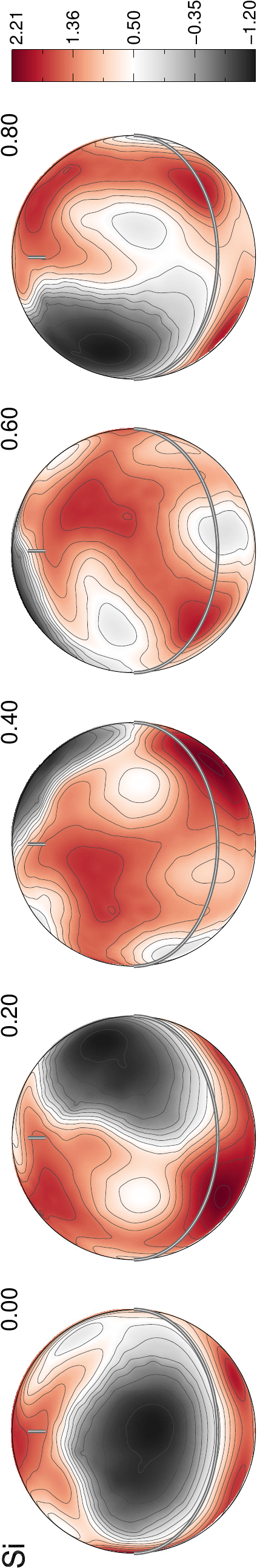}\\\vspace*{0.5cm}
\firrps{15cm}{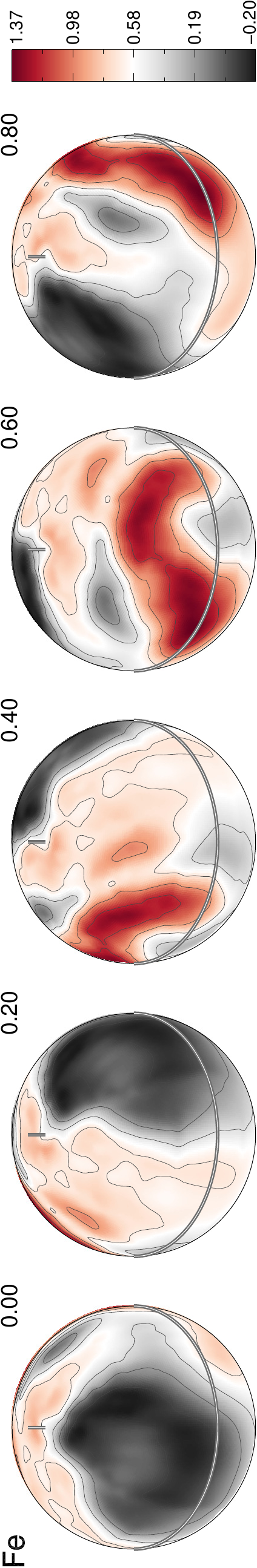}
\caption{Silicon and iron surface abundance distribution reconstructed from the LSD profiles of these elements simultaneously with magnetic field mapping. The star is shown at five rotational phases, indicated next to each plot. The contours over spherical maps are plotted with a 0.25~dex step. The side bars give element abundances in logarithmic units relative to the Sun, [El]\,$\equiv\log (N_{\rm el}/N_{\rm tot})_\star - \log (N_{\rm el}/N_{\rm tot})_\odot$.}
\label{fig:abn}
\end{figure*}

Finally, Fig.~\ref{fig:abn} presents the surface distributions of Si and Fe abundances reconstructed from the LSD profiles of these elements. The Si abundance map spans the range from $+2.2$ to $-1.2$~dex relative to the solar abundance. The Fe map shows a narrower range from $+1.4$ to $-0.2$~dex. Distribution of both elements is dominated by a large spot of relative abundance depletion coinciding with the positive field region, characterised by a weaker predominantly radial field. On the other hand, an overabundance of both elements is correlated with the stronger negative field region and the neighbouring strong positive field region. As one can see from the horizontal field maps in Figs.~\ref{fig:fld_si} and \ref{fig:fld_fe}, this is also the only part of the stellar surface where significant horizontal magnetic field is found. The main Fe overabundance zone is extended in longitude, tracing an arc close to the negative magnetic pole. A similar feature, but spread over a larger area, is visible in the Si abundance map.

\section{Discussion}
\label{discussion}

\subsection{Comparison with previous DI studies}

Previously conventional abundance DI mapping was carried out for \vir\ by \citet{hatzes:1997} and \citet{kuschnig:1999}. The first study presented a Si equivalent width map derived from a single line; the second study obtained abundance maps for five different elements, including Si and Fe for which multiple lines were used. Both analyses gave a high weight to the strongest \ion{Si}{ii} lines. Consequently, their results may be distorted by an interplay of horizontal and vertical abundance inhomogeneities and therefore are less representative than our chemical spot maps derived from the average line profiles. In addition, previous DI studies completely neglected magnetic field and ignored local modifications of the atmospheric structure.

Despite these differences, main features in our Si and Fe abundance maps agree reasonably well with the distributions obtained by \citet{kuschnig:1999}. Taking into account the 0.5 phase offset between the ephemeris adopted in the two studies, we both find an underabundance area around phase 0.0, with Fe underabundance zone trailing slightly behind the one for Si. Also, similar to our study, the maps by \citet{kuschnig:1999} exhibit arch-like features around phase 0.5--0.6 when the Si and Fe abundance reaches maximum. Kuschnig et al. reported a surface abundance variation from +2.5 to $-1.5$ for Si and from +1.3 to $-0.7$ for Fe. These ranges are somewhat larger compared to our study, which may be related to a lower inclination angle adopted by these authors and/or their neglect of the continuum intensity variation due to spots.

\subsection{Verification of the Si abundance map}

\begin{figure*}[!th]
\centering
\fifps{15cm}{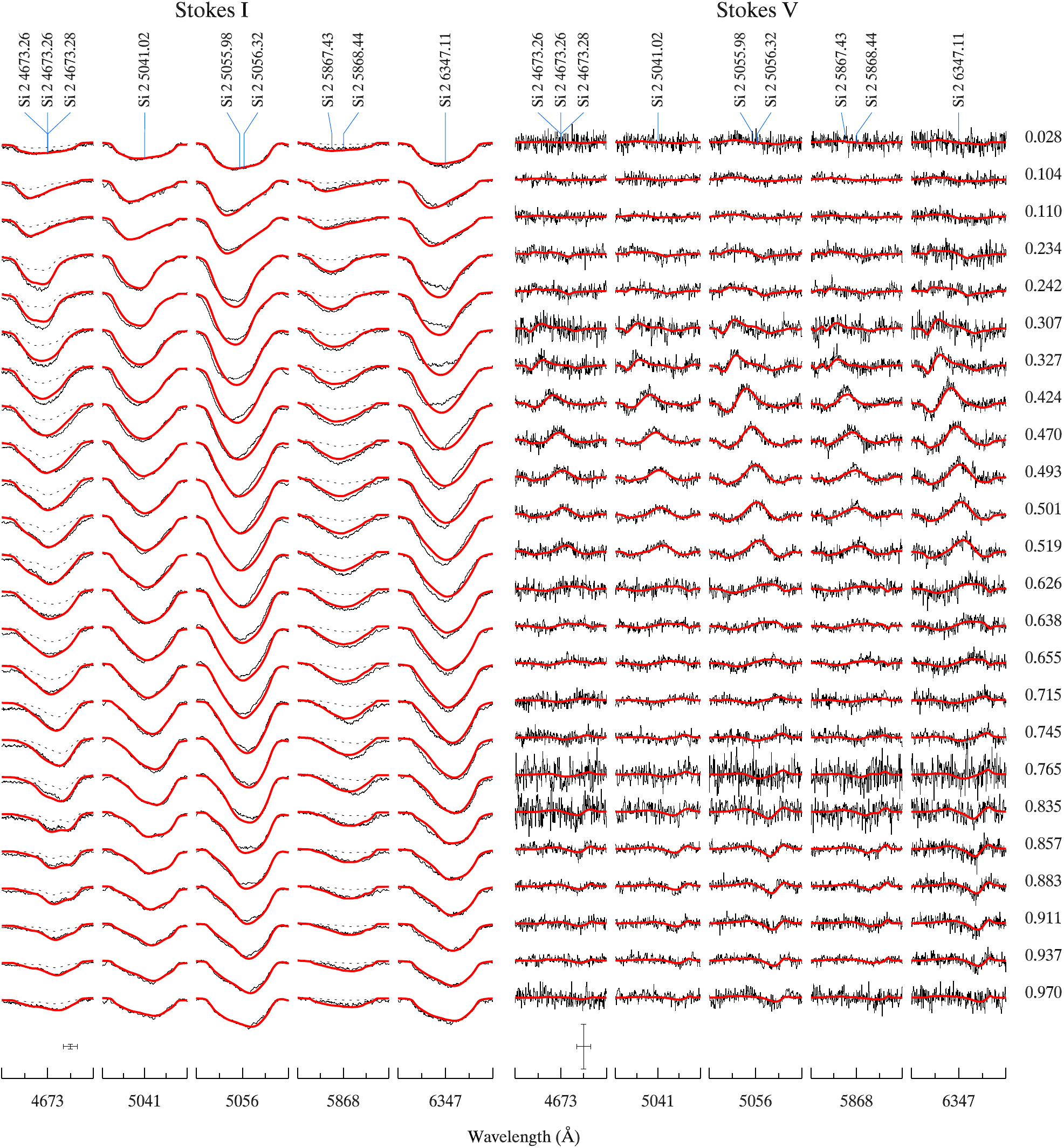}
\caption{Comparison of the observed Stokes $I$ and $V$ profiles (thin solid line) of individual \ion{Si}{ii} lines with the theoretical spectra (thick solid lines) predicted by the magnetic field model and Si abundance distribution derived from silicon LSD profiles (see Figs.~\ref{fig:fld_si} and \ref{fig:abn}). The oscillator strengths of the high-excitation Si blends at $\lambda$ 4673 and 5868~\AA\ have been increased by 0.5--1.5~dex relative to the values given in VALD to match the observations. The spectra predicted with original oscillator strengths are shown with dotted lines. The spectra corresponding to different rotation phases (indicated to the right of the Stokes $V$ panel) are offset vertically. The bars at the lower left corner of each panel indicate the horizontal (1~\AA) and vertical (1\% of $I_{\rm c}$) scales.}
\label{fig:lines}
\end{figure*}

In this study we attempted to derive chemical abundance distributions from LSD profiles, which represent an average over many spectral features with different characteristics. Since this approach is used here for the first time, it is of interest to check how the resulting chemical spot maps reproduce the Stokes $I$ and $V$ profiles of individual spectral lines. In the particular case of \vir\ such an assessment is best carried out for Si since there are many reasonably unblended lines of this element in the stellar spectrum. Using the results of magnetic and chemical abundance mapping with the Si LSD profiles (see Figs.~\ref{fig:fld_si} and \ref{fig:abn}), we computed about a dozen of \ion{Si}{ii} spectral lines and compared their predicted profiles with observations. Figure~\ref{fig:lines} illustrates this comparison for five representative spectral features.

We find that the LSD Si abundance map reproduces fairly well weak and intermediate strength \ion{Si}{ii} lines with the excitation potential of the lower level $E_{\rm i}\approx10$~eV (e.g. \ion{Si}{ii} 5041, 5056~\AA). However, a number of discrepancies is found for other lines. For example, the higher excitation \ion{Si}{ii} lines, such as 4276 and 5268~\AA\ ($E_{\rm i}=13$--15~eV), appear substantially stronger in observations than in our spectrum synthesis. We had to increase their oscillator strengths by 0.5--1.5~dex relative to the values given in VALD to match the observed profiles. On the other hand, very strong silicon lines such as \ion{Si}{ii} 6347 and 6371~\AA\ (the latter not shown in Fig.~\ref{fig:lines}) exhibit a somewhat different profile shape, especially in the phase interval 0.23--0.42 when a distinct flattening of the line cores is observed.

This line to line scatter is almost certainly caused by an inhomogeneous vertical distribution of Si in the atmosphere of \vir. If the Si abundance decreases with height, as found for magnetic CP stars of similar \teff\ \citep{glagolevskii:2005,ryabchikova:2006} and predicted by theoretical radiative diffusion models \citep{leblanc:2009,alecian:2010}, one might expect anomalously strong high-excitation lines (formed in deep atmospheric layers) and profile distortions in strong lower excitation features (sensitive to a large range of heights, including high atmospheric layers). This qualitative scenario describes reasonably well our observations of \vir. This star appears to be an interesting target for studying how the vertical stratification of Si changes across the stellar surface.

In conclusion, our LSD-based Si abundance map provides a satisfactory description of the intensity and circular polarisation spectra of typical individual \ion{Si}{ii}. It does not, however, explain the profiles of certain features likely to be affected by chemical stratification. 

\subsection{Magnetosphere and radio emission of \vir}

\vir\ shows a complex and interesting behaviour in the radio domain. Similar to some other magnetic chemically peculiar stars it shows a smoothly varying, rotationally modulated gyrosynchrotron emission \citep{leone:1996}. In addition, unlike any other known radio-loud stellar source, it exhibits two 100 per cent circularly polarised radio pulses every rotation period \citep{trigilio:2000}. These radio pulses are attributed to the electron cyclotron maser (ECM) mechanism, which produces emission directed at a certain narrow ranges of angles to the magnetic field lines. This emission is believed to be produced in the stellar magnetosphere at a distance of about two stellar radii in the vicinity of one of the magnetic poles. Different modifications of the ECM model have been examined for \vir\ \citep[see][]{lo:2012}, but it is still not entirely clear which one is the most appropriate.

Our work has provided several important constraints for the nature of the magnetosphere of \vir\ and its relation to the unique coherent radio emission of this star. First, our magnetic observations are essentially concurrent with the recent radio studies. This allows us to establish the phase relation between the longitudinal magnetic field curve and the times of arrival of the radio pulses (see Sect.~\ref{longit}) with a much higher certainty than in the previous studies, which had to rely on the \bz\ measurements by \citet{borra:1980} acquired several decades ago. Along these lines, we report new precise phase offsets between the radio pulses at 1.4~GHz and the phases of the zero longitudinal magnetic field.

Furthermore, we found for \vir\ non-negligible deviations from a dipolar magnetic field topology. All previous radio emission models considered purely dipolar field configurations, occasionally speculating about the presence of non-dipolar field components to explain the asymmetry of the radio pulses \citep[e.g.][]{trigilio:2000}. Our ZDI maps characterised these non-dipolar field components and provided detailed information about the surface magnetic field topology that can be used for realistic modelling of the stellar magnetosphere and the radio emission processes. Although such analysis is beyond the scope of our study, it is already clear that our magnetic maps are generally compatible with the asymmetry of the beamed radio emission. Evidently, the negative field region is more compact, is geometrically much better defined and has a stronger field compared to the positive field zone. Therefore, association of the radio pulses with only the negative surface field extremum comes as no surprise given our ZDI results.

Magnetospheric models invoked to explain the radio emission of \vir\ estimated that magnetic field channels the mass loss out to the Alfv\'en radius of $\sim$\,15\,$R$ \citep{trigilio:2004}. This is much larger than the Keplerian radius, thus providing conditions for accumulation of material in corotating clouds or in a warped disk \citep{townsend:2005a}. For \vir\ several studies suggested the presence of a torus of relatively cold, dense material surrounding the star at the magnetic equatorial plane \citep{trigilio:2004,leto:2006}. \citet{lo:2012} argued that refraction from this cold torus represents the most promising version of the ECM mechanism to explain the frequency dependence and shapes of the radio pulse profiles. The circumstellar material in the torus may produce characteristic spectroscopic variation in the hydrogen Balmer lines, similar to the behaviour commonly observed in hotter early-B magnetic stars \citep{townsend:2005,petit:2013}. We searched for such variations using the Narval spectra of \vir. The H$\beta$--$\delta$ lines exhibit smooth changes in their Stark wings due to a non-uniform He distribution over the stellar surface \citep{kuschnig:1999,shulyak:2004}. The maximum line width is found at phase $\approx$\,0.10; the minimum occurs at phase $\approx$\,0.65. In addition, H$\alpha$ shows subtle variation, primarily on the red side of the line core, similar to the spectroscopic changes described earlier by \citet{shore:2004}. Relative to the mean profile the line core alternates between a lack of absorption at phase $\approx$\,0.24 and an excess at phase $\approx$\,0.75. This variability in H$\alpha$ is, again, likely a result of  He abundance spots rather than cold, dense gas trapped in the stellar magnetosphere. Thus, our observations provide no direct evidence for the existence of circumstellar material in the vicinity of \vir.

\subsection{Conclusions}

In this study we developed a new Zeeman Doppler imaging procedure, combining for the first time the LSD multi-line technique with the detailed magnetic spectrum synthesis using realistic model atmospheres. This new magnetic inversion methodology was successfully applied to a set of high-quality circular polarisation observations of the well-known magnetic chemically peculiar star \vir. The main conclusion of our investigation can be summarised as follows.
\begin{itemize}
\item The new ZDI methodology enables simultaneous reconstruction of the stellar magnetic field topology and distribution of chemical abundance (or temperature for cool active stars) from the LSD Stokes profiles without making simplifying approximations typical of previous ZDI studies. In particular, the new ZDI method does not assume that a response of the LSD profiles to the variation of magnetic field, chemical abundance or temperature is equivalent to that of a single spectral line.
\item The revised fundamental parameters of \vir\ suggest that this is a young magnetic star observed from an intermediate inclination. It is a relatively rapid rotator, presumably due to its youth, but its equatorial rotational velocity is not high enough to significantly distort the spherical shape of the stellar surface.
\item The magnetic field topology of \vir\ is generally dipolar-like, but is characterised by a strong field modulus asymmetry between the two poles and by a major deviation from an axisymmetric configuration. This field geometry is mainly poloidal and nearly all field energy is contained in the harmonic modes with the angular degree $\ell$\,=\,1--3.
\item Our calculations prove that neither the purely dipolar magnetic field model nor a more complex dipole plus quadrupole field topology derived from the longitudinal field curve of \vir\ can describe the observed circular polarisation signatures in spectral lines. 
\item The chemical abundance maps of Si and Fe are dominated by a large underabundance region where the field is positive, predominantly radial, and weak. In contrast, the spots of element overabundance occur at the stronger negative and positive field regions where a significant horizontal field component is also present.
\item The magnetic and chemical abundance maps obtained from the Si LSD profiles match variability of the Stokes $I$ and $V$ spectra in individual \ion{Si}{ii} lines. However, observations show some line to line scatter in the Si profile shapes and intensities, which we tentatively attribute to a vertical stratification of silicon in the atmosphere of \vir.
\item The magnetic field maps derived in our study explain the asymmetry of the pulsed radio emission of \vir\ and provide detailed constraints for future magnetospheric modelling.
\end{itemize}

\begin{acknowledgements}
OK is a Royal Swedish Academy of Sciences Research Fellow, supported by the grants from the Knut and Alice Wallenberg Foundation, Swedish Research Council, and the G\"oran Gustafsson Foundation. 
TL acknowledges the support by the FWF NFN project S11601-N16 ``Pathways to Habitability: From Disks to Active Stars, Planets and Life'', and the related FWF NFN subproject S116 604-N16, as well as the financial contributions of the Austrian Agency for International Cooperation in Education and Research (CZ-10/2012). 
EA, CN, and the MiMeS collaboration acknowledge financial support from the Programme National de Physique Stellaire (PNPS) of INSU/CNRS.
The computations presented in this paper were performed on resources provided by SNIC through Uppsala Multidisciplinary Center for Advanced Computational Science (UPPMAX) under project snic2013-11-24.
\end{acknowledgements}


\begin{thebibliography}{60}
\expandafter\ifx\csname natexlab\endcsname\relax\def\natexlab#1{#1}\fi

\bibitem[{{Alecian} \& {Stift}(2010)}]{alecian:2010}
{Alecian}, G. \& {Stift}, M.~J. 2010, \aap, 516, A53

\bibitem[{{Alecian} \& {Vauclair}(1981)}]{alecian:1981}
{Alecian}, G. \& {Vauclair}, S. 1981, \aap, 101, 16

\bibitem[{{Asplund} {et~al.}(2009){Asplund}, {Grevesse}, {Sauval}, \&
  {Scott}}]{asplund:2009}
{Asplund}, M., {Grevesse}, N., {Sauval}, A.~J., \& {Scott}, P. 2009, \araa, 47,
  481

\bibitem[{{Bagnulo} {et~al.}(2009){Bagnulo}, {Landolfi}, {Landstreet}, {Landi
  Degl'Innocenti}, {Fossati}, \& {Sterzik}}]{bagnulo:2009}
{Bagnulo}, S., {Landolfi}, M., {Landstreet}, J.~D., {et~al.} 2009, \pasp, 121,
  993

\bibitem[{{Bertelli} {et~al.}(2009){Bertelli}, {Nasi}, {Girardi}, \&
  {Marigo}}]{bertelli:2009}
{Bertelli}, G., {Nasi}, E., {Girardi}, L., \& {Marigo}, P. 2009, \aap, 508, 355

\bibitem[{{Borra} \& {Landstreet}(1980)}]{borra:1980}
{Borra}, E.~F. \& {Landstreet}, J.~D. 1980, \apjs, 42, 421

\bibitem[{{Collins}(1963)}]{collins:1963}
{Collins}, II, G.~W. 1963, \apj, 138, 1134

\bibitem[{{Deutsch}(1952)}]{deutsch:1952}
{Deutsch}, A.~J. 1952, \apj, 116, 536

\bibitem[{{Donati} {et~al.}(1997){Donati}, {Semel}, {Carter}, {Rees}, \&
  {Collier Cameron}}]{donati:1997}
{Donati}, J.-F., {Semel}, M., {Carter}, B.~D., {Rees}, D.~E., \& {Collier
  Cameron}, A. 1997, \mnras, 291, 658

\bibitem[{{Donati} {et~al.}(2006){Donati}, {Howarth}, {Jardine}, {Petit},
  {Catala}, {Landstreet}, {Bouret}, {Alecian}, {Barnes}, {Forveille},
  {Paletou}, \& {Manset}}]{donati:2006b}
{Donati}, J.-F., {Howarth}, I.~D., {Jardine}, M.~M., {et~al.} 2006, \mnras,
  370, 629

\bibitem[{{Folsom} {et~al.}(2008){Folsom}, {Wade}, {Kochukhov}, {Alecian},
  {Catala}, {Bagnulo}, {B{\"o}hm}, {Bouret}, {Donati}, {Grunhut}, {Hanes}, \&
  {Landstreet}}]{folsom:2008}
{Folsom}, C.~P., {Wade}, G.~A., {Kochukhov}, O., {et~al.} 2008, \mnras, 391,
  901

\bibitem[{{Glagolevskii} {et~al.}(2005){Glagolevskii}, {Ryabchikova}, \&
  {Chountonov}}]{glagolevskii:2005}
{Glagolevskii}, Y.~V., {Ryabchikova}, T.~A., \& {Chountonov}, G.~A. 2005,
  Astronomy Letters, 31, 327

\bibitem[{{Glagolevskij} \& {Gerth}(2002)}]{glagolevskij:2002}
{Glagolevskij}, Y.~V. \& {Gerth}, E. 2002, \aap, 382, 935

\bibitem[{{Goncharskii} {et~al.}(1983){Goncharskii}, {Ryabchikova}, {Stepanov},
  {Khokhlova}, \& {Yagola}}]{goncharskii:1983}
{Goncharskii}, A.~V., {Ryabchikova}, T.~A., {Stepanov}, V.~V., {Khokhlova},
  V.~L., \& {Yagola}, A.~G. 1983, Soviet Astronomy, 27, 49

\bibitem[{{Hardie}(1958)}]{hardie:1958}
{Hardie}, R. 1958, \apj, 127, 620

\bibitem[{{Hatzes}(1997)}]{hatzes:1997}
{Hatzes}, A.~P. 1997, \mnras, 288, 153

\bibitem[{{Hiesberger} {et~al.}(1995){Hiesberger}, {Piskunov}, {Bonsack},
  {Weiss}, {Ryabchikova}, \& {Kuschnig}}]{hiesberger:1995}
{Hiesberger}, F., {Piskunov}, N., {Bonsack}, W.~K., {et~al.} 1995, \aap, 296,
  473

\bibitem[{{Kochukhov}(2007)}]{kochukhov:2007d}
{Kochukhov}, O. 2007, in Physics of Magnetic Stars, ed. I.~I. {Romanyuk} \&
  D.~O. {Kudryavtsev}, 109--118

\bibitem[{{Kochukhov} \& {Bagnulo}(2006)}]{kochukhov:2006}
{Kochukhov}, O. \& {Bagnulo}, S. 2006, \aap, 450, 763

\bibitem[{{Kochukhov} \& {Sudnik}(2013)}]{kochukhov:2013b}
{Kochukhov}, O. \& {Sudnik}, N. 2013, \aap, 554, A93

\bibitem[{{Kochukhov} \& {Wade}(2010)}]{kochukhov:2010}
{Kochukhov}, O. \& {Wade}, G.~A. 2010, \aap, 513, A13

\bibitem[{{Kochukhov} {et~al.}(2004){Kochukhov}, {Bagnulo}, {Wade}, {Sangalli},
  {Piskunov}, {Landstreet}, {Petit}, \& {Sigut}}]{kochukhov:2004d}
{Kochukhov}, O., {Bagnulo}, S., {Wade}, G.~A., {et~al.} 2004, \aap, 414, 613

\bibitem[{{Kochukhov} {et~al.}(2010){Kochukhov}, {Makaganiuk}, \&
  {Piskunov}}]{kochukhov:2010a}
{Kochukhov}, O., {Makaganiuk}, V., \& {Piskunov}, N. 2010, \aap, 524, A5

\bibitem[{{Kochukhov} {et~al.}(2012){Kochukhov}, {Wade}, \&
  {Shulyak}}]{kochukhov:2012}
{Kochukhov}, O., {Wade}, G.~A., \& {Shulyak}, D. 2012, \mnras, 421, 3004

\bibitem[{{Kochukhov} {et~al.}(2013){Kochukhov}, {Mantere}, {Hackman}, \&
  {Ilyin}}]{kochukhov:2013}
{Kochukhov}, O., {Mantere}, M.~J., {Hackman}, T., \& {Ilyin}, I. 2013, \aap,
  550, A84

\bibitem[{{Krti{\v c}ka} {et~al.}(2012){Krti{\v c}ka}, {Mikul{\'a}{\v s}ek},
  {L{\"u}ftinger}, {Shulyak}, {Zverko}, {{\v Z}i{\v z}{\v n}ovsk{\'y}}, \&
  {Sokolov}}]{krticka:2012}
{Krti{\v c}ka}, J., {Mikul{\'a}{\v s}ek}, Z., {L{\"u}ftinger}, T., {et~al.}
  2012, \aap, 537, A14

\bibitem[{{Kupka} {et~al.}(1999){Kupka}, {Piskunov}, {Ryabchikova}, {Stempels},
  \& {Weiss}}]{kupka:1999}
{Kupka}, F., {Piskunov}, N., {Ryabchikova}, T.~A., {Stempels}, H.~C., \&
  {Weiss}, W.~W. 1999, \aaps, 138, 119

\bibitem[{{Kuschnig} {et~al.}(1999){Kuschnig}, {Ryabchikova}, {Piskunov},
  {Weiss}, \& {Gelbmann}}]{kuschnig:1999}
{Kuschnig}, R., {Ryabchikova}, T.~A., {Piskunov}, N.~E., {Weiss}, W.~W., \&
  {Gelbmann}, M.~J. 1999, \aap, 348, 924

\bibitem[{{Landstreet} \& {Borra}(1977)}]{landstreet:1977}
{Landstreet}, J.~D. \& {Borra}, E.~F. 1977, \apjl, 212, L43

\bibitem[{{Landstreet} \& {Mathys}(2000)}]{landstreet:2000}
{Landstreet}, J.~D. \& {Mathys}, G. 2000, \aap, 359, 213

\bibitem[{{LeBlanc} {et~al.}(2009){LeBlanc}, {Monin}, {Hui-Bon-Hoa}, \&
  {Hauschildt}}]{leblanc:2009}
{LeBlanc}, F., {Monin}, D., {Hui-Bon-Hoa}, A., \& {Hauschildt}, P.~H. 2009,
  \aap, 495, 937

\bibitem[{{Leone} {et~al.}(1996){Leone}, {Umana}, \& {Trigilio}}]{leone:1996}
{Leone}, F., {Umana}, G., \& {Trigilio}, C. 1996, \aap, 310, 271

\bibitem[{{Leto} {et~al.}(2006){Leto}, {Trigilio}, {Buemi}, {Umana}, \&
  {Leone}}]{leto:2006}
{Leto}, P., {Trigilio}, C., {Buemi}, C.~S., {Umana}, G., \& {Leone}, F. 2006,
  \aap, 458, 831

\bibitem[{{Lipski} \& {St{\c e}pie{\'n}}(2008)}]{lipski:2008}
{Lipski}, {\L}. \& {St{\c e}pie{\'n}}, K. 2008, \mnras, 385, 481

\bibitem[{{Lo} {et~al.}(2012){Lo}, {Bray}, {Hobbs}, {Murphy}, {Gaensler},
  {Melrose}, {Ravi}, {Manchester}, \& {Keith}}]{lo:2012}
{Lo}, K.~K., {Bray}, J.~D., {Hobbs}, G., {et~al.} 2012, \mnras, 421, 3316

\bibitem[{{Marsden} {et~al.}(2011){Marsden}, {Jardine}, {Ram{\'{\i}}rez
  V{\'e}lez}, {Alecian}, {Brown}, {Carter}, {Donati}, {Dunstone}, {Hart},
  {Semel}, \& {Waite}}]{marsden:2011}
{Marsden}, S.~C., {Jardine}, M.~M., {Ram{\'{\i}}rez V{\'e}lez}, J.~C., {et~al.}
  2011, \mnras, 413, 1922

\bibitem[{{Michaud} {et~al.}(1981){Michaud}, {Charland}, \&
  {Megessier}}]{michaud:1981}
{Michaud}, G., {Charland}, Y., \& {Megessier}, C. 1981, \aap, 103, 244

\bibitem[{{Mikul{\'a}{\v s}ek} {et~al.}(2011){Mikul{\'a}{\v s}ek}, {Krti{\v
  c}ka}, {Henry}, {Jan{\'{\i}}k}, {Zverko}, {{\v Z}i{\v z}{\v n}ovsk{\'y}},
  {Zejda}, {Li{\v s}ka}, {Zv{\v e}{\v r}ina}, {Kudrjavtsev}, {Romanyuk},
  {Sokolov}, {L{\"u}ftinger}, {Trigilio}, {Neiner}, \& {de
  Villiers}}]{mikulasek:2011}
{Mikul{\'a}{\v s}ek}, Z., {Krti{\v c}ka}, J., {Henry}, G.~W., {et~al.} 2011,
  \aap, 534, L5

\bibitem[{{Morin} {et~al.}(2008){Morin}, {Donati}, {Petit}, {Delfosse},
  {Forveille}, {Albert}, {Auri{\`e}re}, {Cabanac}, {Dintrans}, {Fares},
  {Gastine}, {Jardine}, {Ligni{\`e}res}, {Paletou}, {Ramirez Velez}, \&
  {Th{\'e}ado}}]{morin:2008}
{Morin}, J., {Donati}, J.-F., {Petit}, P., {et~al.} 2008, \mnras, 390, 567

\bibitem[{{Netopil} {et~al.}(2008){Netopil}, {Paunzen}, {Maitzen}, {North}, \&
  {Hubrig}}]{netopil:2008}
{Netopil}, M., {Paunzen}, E., {Maitzen}, H.~M., {North}, P., \& {Hubrig}, S.
  2008, \aap, 491, 545

\bibitem[{{North}(1998)}]{north:1998a}
{North}, P. 1998, \aap, 334, 181

\bibitem[{{Petit} {et~al.}(2004){Petit}, {Donati}, {Wade}, {Landstreet},
  {Bagnulo}, {L{\"u}ftinger}, {Sigut}, {Shorlin}, {Strasser}, {Auri{\`e}re}, \&
  {Oliveira}}]{petit:2004a}
{Petit}, P., {Donati}, J., {Wade}, G.~A., {et~al.} 2004, \mnras, 348, 1175

\bibitem[{{Petit} {et~al.}(2013){Petit}, {Owocki}, {Wade}, {Cohen},
  {Sundqvist}, {Gagn{\'e}}, {Ma{\'{\i}}z Apell{\'a}niz}, {Oksala}, {Bohlender},
  {Rivinius}, {Henrichs}, {Alecian}, {Townsend}, {ud-Doula}, \& {MiMeS
  Collaboration}}]{petit:2013}
{Petit}, V., {Owocki}, S.~P., {Wade}, G.~A., {et~al.} 2013, \mnras, 429, 398

\bibitem[{{Piskunov} \& {Kochukhov}(2002)}]{piskunov:2002a}
{Piskunov}, N. \& {Kochukhov}, O. 2002, \aap, 381, 736

\bibitem[{{Piskunov} {et~al.}(1995){Piskunov}, {Kupka}, {Ryabchikova}, {Weiss},
  \& {Jeffery}}]{piskunov:1995}
{Piskunov}, N.~E., {Kupka}, F., {Ryabchikova}, T.~A., {Weiss}, W.~W., \&
  {Jeffery}, C.~S. 1995, \aaps, 112, 525

\bibitem[{{Pyper} {et~al.}(1998){Pyper}, {Ryabchikova}, {Malanushenko},
  {Kuschnig}, {Plachinda}, \& {Savanov}}]{pyper:1998}
{Pyper}, D.~M., {Ryabchikova}, T., {Malanushenko}, V., {et~al.} 1998, \aap,
  339, 822

\bibitem[{{Pyper} {et~al.}(2013){Pyper}, {Stevens}, \& {Adelman}}]{pyper:2013}
{Pyper}, D.~M., {Stevens}, I.~R., \& {Adelman}, S.~J. 2013, \mnras, 431, 2106

\bibitem[{{Ravi} {et~al.}(2010){Ravi}, {Hobbs}, {Wickramasinghe}, {Champion},
  {Keith}, {Manchester}, {Norris}, {Bray}, {Ferrario}, \&
  {Melrose}}]{ravi:2010}
{Ravi}, V., {Hobbs}, G., {Wickramasinghe}, D., {et~al.} 2010, \mnras, 408, L99

\bibitem[{{Ryabchikova} {et~al.}(2006){Ryabchikova}, {Ryabtsev}, {Kochukhov},
  \& {Bagnulo}}]{ryabchikova:2006}
{Ryabchikova}, T., {Ryabtsev}, A., {Kochukhov}, O., \& {Bagnulo}, S. 2006,
  \aap, 456, 329

\bibitem[{{Shore} {et~al.}(2004){Shore}, {Bohlender}, {Bolton}, {North}, \&
  {Hill}}]{shore:2004}
{Shore}, S.~N., {Bohlender}, D.~A., {Bolton}, C.~T., {North}, P., \& {Hill},
  G.~M. 2004, \aap, 421, 203

\bibitem[{{Shulyak} {et~al.}(2004){Shulyak}, {Tsymbal}, {Ryabchikova},
  {St{\"u}tz}, \& {Weiss}}]{shulyak:2004}
{Shulyak}, D., {Tsymbal}, V., {Ryabchikova}, T., {St{\"u}tz}, C., \& {Weiss},
  W.~W. 2004, \aap, 428, 993

\bibitem[{{Tikhonov} \& {Arsenin}(1977)}]{tikhonov:1977}
{Tikhonov}, A.~N. \& {Arsenin}, V.~Y. 1977, {Solution of Ill-posed Problems}
  (Wiley: New York)

\bibitem[{{Townsend} \& {Owocki}(2005)}]{townsend:2005a}
{Townsend}, R.~H.~D. \& {Owocki}, S.~P. 2005, \mnras, 357, 251

\bibitem[{{Townsend} {et~al.}(2005){Townsend}, {Owocki}, \&
  {Groote}}]{townsend:2005}
{Townsend}, R.~H.~D., {Owocki}, S.~P., \& {Groote}, D. 2005, \apjl, 630, L81

\bibitem[{{Trigilio} {et~al.}(2000){Trigilio}, {Leto}, {Leone}, {Umana}, \&
  {Buemi}}]{trigilio:2000}
{Trigilio}, C., {Leto}, P., {Leone}, F., {Umana}, G., \& {Buemi}, C. 2000,
  \aap, 362, 281

\bibitem[{{Trigilio} {et~al.}(2004){Trigilio}, {Leto}, {Umana}, {Leone}, \&
  {Buemi}}]{trigilio:2004}
{Trigilio}, C., {Leto}, P., {Umana}, G., {Leone}, F., \& {Buemi}, C.~S. 2004,
  \aap, 418, 593

\bibitem[{{Trigilio} {et~al.}(2008){Trigilio}, {Leto}, {Umana}, {Buemi}, \&
  {Leone}}]{trigilio:2008}
{Trigilio}, C., {Leto}, P., {Umana}, G., {Buemi}, C.~S., \& {Leone}, F. 2008,
  \mnras, 384, 1437

\bibitem[{{Trigilio} {et~al.}(2011){Trigilio}, {Leto}, {Umana}, {Buemi}, \&
  {Leone}}]{trigilio:2011}
{Trigilio}, C., {Leto}, P., {Umana}, G., {Buemi}, C.~S., \& {Leone}, F. 2011,
  \apjl, 739, L10

\bibitem[{{Ud-Doula} {et~al.}(2009){Ud-Doula}, {Owocki}, \&
  {Townsend}}]{ud-doula:2009}
{Ud-Doula}, A., {Owocki}, S.~P., \& {Townsend}, R.~H.~D. 2009, \mnras, 392,
  1022

\bibitem[{{van Leeuwen}(2007)}]{van-leeuwen:2007}
{van Leeuwen}, F. 2007, \aap, 474, 653

\end{thebibliography}

\end{document}